# Energy Efficient Placement of Workloads in Composable Data Center Networks

Opeyemi O. Ajibola, Taisir E. H. El-Gorashi, and Jaafar M. H. Elmirghani, *Fellow, IEEE*

*Abstract*—**This paper studies the energy efficiency of composable datacentre (DC) infrastructures over network topologies. Using a mixed integer linear programming (MILP) model, we compare the performance of disaggregation at rack-scale and pod-scale over selected electrical, optical and hybrid network topologies relative to a traditional DC. Relative to a pod-scale DC, the results show that physical disaggregation at rack-scale is sufficient for optimal efficiency when the optical network topology is adopted and resource components are allocated in a suitable manner. The optical network topology also enables optimal energy efficiency in composable DCs. The paper also studies logical disaggregation of traditional DC servers over an optical network topology. Relative to physical disaggregation at rack-scale, logical disaggregation of server resources within each rack enables marginal fall in the total DC power consumption (TDPC) due to improved resource demands placement. Hence, an adaptable composable infrastructure that can support both in memory (access) latency sensitive and insensitive workloads is enabled. We also conduct a study of the adoption of micro-service architecture in both traditional and composable DCs. Our results show that increasing the modularity of workloads improves the energy efficiency in traditional DCs, but disproportionate utilization of DC resources persists. A combination of disaggregation and micro-services achieved up to 23% reduction in the TDPC of the traditional DC by enabling optimal resources utilization and energy efficiencies. Finally, we propose a heuristic for energy efficient placement of workloads in composable DCs which replicates the trends produced by the MILP model formulated in this paper.**

*Index Terms*—**Composable infrastructures, energy efficient data centres, MILP, micro-services, optical networks, rack-scale, software defined infrastructures.**

## I. INTRODUCTION

DATACENTERS are critical infrastructures that provide platforms driving wide adoption of digital technologies. These indispensable infrastructures provide computing resources needed to run public internet-facing applications and private enterprise-critical applications alike in environments that support the requirements of cloud computing and data analytics applications. The requirements of cloud computing and data analytic applications include on-demand resource provisioning, multitenant isolation, parallel computation, and security. Examples of such applications include web services, web-search, instant messaging and social media, distributed file systems, analytics, and content delivery applications. In addition to these widely used applications, strong emergence of disruptive technologies such as network function virtualization, the Internet of things (IoT), artificial intelligence, big data analytics and smart grids and cities, is also expected to increase the variety of applications deployed in datacenters (DCs) [1] in the near future.

DCs comprise of compute, storage and network resources and peripheral components (including management and orchestration platforms, power and cooling systems which are critical to daily operation and management of DCs). In traditional DCs, a server is the basic unit from which warehouse scale DCs are composed. This server is a modular node with small quantities of in-built compute, storage and network resources. Lately, storage resource in DCs are aggregated into storage systems such as storage area network (SAN) and network attached storage (NAS). Up to 48 servers are placed in cabinet-like structures called racks as shown in Fig. 1, servers within a rack are inter-connected by an intra-rack communication network. Multiple co-located racks are connected by an inter-rack network to form a DC site/cluster. DCs have been classified into university, private enterprise and cloud categories based on the mix of applications in the DC and the organization that own and operate the infrastructure [2]. While university campus and private enterprise DCs are on premise infrastructures with a few hundreds of servers, cloud DCs usually comprise of tens of thousands of servers which may be distributed across the globe to ensure quality of service and regulation requirements are met.

In recent times, expected growth in the proliferation of DCs and in the number of applications running in their underlying infrastructures are motivations for increased flexibility, resource utilization efficiency and energy efficiency of DCs to attain desired performance at low cost and low power consumption. However, the traditional DC infrastructure has known shortfalls. These shortfalls include integration of resource and server proportionalities during infrastructure rollouts and upgrades, resource fragmentation and utilization inefficiencies [3], [4], high workload blockage probability [4], [5], high infrastructure capital expenditure (CAPEX) and

This work was supported in part by the Engineering and Physical Sciences Research Council (EPSRC), in part by INTelligent Energy aware NETworks (INTERNET) under Grant EP/H040536/1, in part by SwiTching And tRansmission (STAR) under Grant EP/K016873/1, and in part by Terabit Bidirectional Multi-user Optical Wireless System (TOWS) project under Grant EP/S016570/1. The first author would like to acknowledge his PhD scholarship awarded by the Petroleum Technology Trust Fund (PTDF), Nigeria. All data is provided in the results section of this paper.

The authors are with the School of Electronic and Electrical Engineering, University of Leeds, Leeds, LS2 9JT, U.K. (e-mail: el14oa@leeds.ac.uk; t.e.h.elgorashi@leeds.ac.uk; j.m.h.elmirghani@leeds.ac.uk).

operational expenditure (OPEX) [6], [7] and poor support for wide range of emerging applications (such as applications adopting in-memory computing architecture) that require ultra-low latency access to large data sets. The rigid architecture of traditional DCs also limits the integration/adoption of advanced hardware technologies such as emerging storage class memory (SCM)[8] with higher storage capacity and lower latency compared to traditional HDDs. To address these limitations, concepts such as server hardware disaggregation and software definition of DC resources have become prevalent in both academic and industry led DC infrastructure discourses. In addition, the closing throughput gap between backplane and traditional networking technologies [7] such as peripheral component interconnect express (PCIe) and Ethernet is motivating the adventure into novel DC infrastructures.

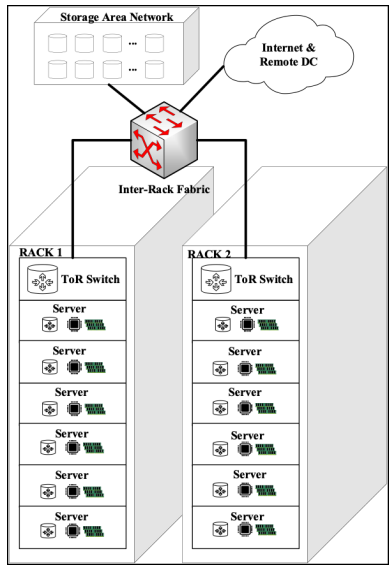

Fig. 1. Traditional data center infrastructure. The figure shows two racks in a traditional DC comprised of servers and a storage area network.

Software definition of DC resources proposes the virtualization of compute, storage, and network resources in DCs for dynamic provisioning and configuration via software. Thus, enabling increased flexibility and agility of DCs for temporal support of diverse workloads while reusing the same physical resources. On the other hand, server hardware disaggregation proposes the logical or physical separation of server resources into pools of homogenous resource, from which resources can be temporally allocated to workloads on-demand. These two concepts are geared towards ensuring reduction in CAPEX and OPEX of DCs while making optimal use of available resources to meet workloads resource demands. Software definition of DC resource and server hardware disaggregation concepts are critical technologies required for composable DC infrastructures to emerge.

Several recent studies have focused on gains as a result of hardware disaggregation in a software defined environment from perspectives such as resource utilization, workload acceptance, consolidation [5], [9], energy efficiency and power consumption [4], [6]. However, the optimal scale of resource disaggregation remains an open issue. In this paper, we compare different scales of resource disaggregation in composable DC infrastructures via the formulation of a mixed integer linear programming (MILP) model. We also review electrical, optical and hybrid network topologies proposed for composable DC infrastructure in existing literature and evaluate the performance of selected prototype topologies. The concept of logical resource disaggregation in composable DC where resource utilization boundaries are virtually relaxed is also explored. Finally, we evaluate the impact of using micro-services to form integrated workloads in both traditional and composable DCs over the deployment of monolithic workloads. The results of comparisons in this paper are expected to aid university, enterprises, telecommunication companies and cloud DC operators to determine the right DC infrastructure to deploy for optimal resource utilization with minimal CAPEX and OPEX. The results will also aid the design of network topologies for composable DC infrastructures.

The remainder of this paper is outlined as follows; in Section II, we review composable DC infrastructures, their enabling technologies, and discuss the spectrum of feasible composable DC infrastructure. In Section III, we review network topologies proposed for composable DC infrastructure in recent literatures. In Section IV, we introduce a MILP model to represent and compare disaggregated composable DC infrastructures under selected network topologies. Section V discusses the results obtained from the MILP model defined in Section IV. In Section VI, the impact of micro-services in composable DCs is studied. In Section VII, we propose a heuristic algorithm for the energy efficient placement of workloads in composable infrastructure. Finally, we conclude the paper in Section VIII.

## II. Overview of Composable DC Infrastructures

Composable DC infrastructures require the separation (disaggregation) of traditional server resources into independent compute, storage, and network pools. These disaggregated resources are composed on-demand to form virtual computing systems which can be decomposed and recomposed using a software defined orchestration and management layer via unified application programming interface (API). With this ability a dynamic infrastructure with fluid pool of resources is enabled. This pool of resources can be composed on-demand to support the resource requirements of a unique workload of any type and size for a defined duration. Composable DC infrastructure addresses challenges of traditional DC infrastructures in the following ways:

- Improving resource modularity and lifecycle management;
- Removing the bottlenecks associated with bespoke (workload-specific) DC hardware;
- Improving agility and speed of DC resources composition for legacy and emerging workloads;
- Increasing efficient utilization of DC resources; and
- Improving overall energy and cost efficiencies in DCs.

Disaggregation and software definition of DC resources are primary concepts essential to achieving composable DC infrastructures. Additionally, appropriate network topologies are also required between resources in composable DCs.

### A. Resource Disaggregation and Allocation

Physical disaggregation of server resources can be performed at different scales i.e. rack-scale and pod-scale [7]. At rack-scale, physically disaggregated resources are integrated into nodes with homogenous resources, i.e. nodes of different resource types (CPU, memory, storage, and IO) are allocated to a rack as shown in Fig. 2a to ensure compute system composition within a rack i.e. at the rack-level. The resources within each rack are only available to other co-rack resources. On the other hand, at pod-scale, nodes of the same resource type are allocated to a rack, racks of different resource types are allocated to a pod as shown in Fig. 2b to ensure compute system composition within a pod i.e. at the pod-level. The resources within each rack in a pod are available to other resources within the domain of that pod. Section IV of this paper compares physical disaggregation at rack-scale and at pod-scale to evaluate the optimal level of disaggregation. It is important to note that the concepts of rack and pod scale disaggregation discussed so far represent physical disaggregation of DC resources. It is also possible to disaggregated resources logically. Section IV of this paper also explores a composable DC derived from logical disaggregation of resources of a traditional DC.

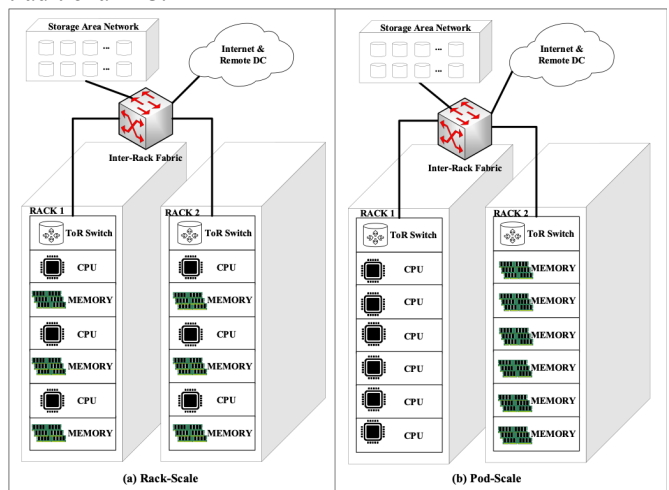

Fig. 2. Rack-scale and pod-scale composable DC infrastructures

Disaggregation of DC server resources at any scale requires communication networks and centralized resource orchestration and management to enable dynamic provisioning of disaggregated resources while sustaining the performance observed in traditional DCs. Hence, the design of appropriate communication fabrics to support latency and throughput of communication between disaggregated resources is essential. Such requirements between compute and IO/storage resources are adequately satisfied by network topologies built using Fibre Channel, Ethernet, and InfiniBand networking technologies [10]. Thus, practical systems with disaggregated IO and storage resources exist today. On the other hand, disaggregation of compute into independent CPU and main memory pools is daunting because of the high communication bandwidth and ultra-low latency required between these two resource components [11]. We review some proposed network topologies for composable DC infrastructure in Section III of this paper.

### B. Software Defined Infrastructures

The advent of software defined infrastructures (SDI) provides centralized orchestration, management, and control layer for the composition of disaggregated DC resources. In software defined infrastructures, physical compute, network, and storage resources are abstracted to enable automated workload-centric composition, monitoring and management of DC resources via a centralized software which can dynamically adjust resource composition to maintain workload defined policies. Server, network, and storage virtualizations are three key technologies at the heart of SDI. In addition, the API between the business logic layer and the centralised control plane of SDI enables programmable pools of hardware resources that can be provisioned based on workload demands. This paper does not focus on SDI and its core technologies, rather we assume that this capability is available while we focus on the evaluation of optimal scale of resource disaggregation under effective software definition. Interested readers are referred to [12]–[14] for discussions on SDI and its underlying components.

### C. Related Works

Prior to this work, extensive studies on the energy efficiency of communication networks that support the increasingly growing exchange of data between data centres and end-users due to rising uptake of the digital solutions and services have been conducted in [15]–[19]. In [15], authors formulated linear programming model that optimized the location of DCs over the core network to minimize the power consumed in networks and also studied how to replicate content to minimize power consumption using a linear programming model. This authors also studied the optimal location of DCs in the network when renewable energy sources are available. In [16], the authors proposed the use of renewable energy to reduce the CO2 emission of IP over WDM networks by formulating a linear programming model that minimized energy consumption in networks when renewable energy is used. The optimal location of renewable energy sources in the network was also studied in this work. A study of energy efficient physical topologies for core IP over WDM networks was conducted using MILP model that minimizes the total network power consumption in [17]. The hotspot impact of the presence of DCs in the core network was also considered along with the benefits of adopting a full mesh or star topology for core networks. When renewable energy sources were available, the results showed that full mesh and star topologies increased utilization of renewable energy sources.

In [18], the authors optimized the placement of a range of cloud services, including content delivery services, storage as a service and compute services, over the IP over WDM core network. MILP models were formulated to optimize the placement of such while minimizing power consumption and comparable heuristics were developed using insights from the MILP model results. The authors studied the impact of a popular peer to peer (P2P) protocol (BitTorrent) on the energy efficiency of core IP over WDM networks via the use of a MILP model and a comparable heuristic was proposed in [19]. Additionally, the authors validated the results obtained via simulation empirically and formulated a MILP model to compare the power consumption of delivering video on demand services over IP over WDM networks by using content delivery network (CDN), P2P and hybrid CDN-P2P architectures. In this work, we exit the domain of communication networks that connect the DCs to end-users to focus on the communication networks within DCs. We explicitly focus on networks that can support the physically or logically disaggregated resource components in composable DC infrastructures.

The concept of composable DC infrastructure has been explored by notable vendors in the IT infrastructure industry such as HPE [20] and Cisco [21] with the primary aim of creating a programmable infrastructure that can simultaneously support legacy enterprise applications used to ensure efficient operations of businesses and emerging applications such as mobile, social media, big data and cloud computing applications which are revenue generators for businesses. However, the proposed prototypes adopt partial disaggregation. Hence, CPU and memory resources are integrated into a module called 'compute module' while storage and IO resources are separated and placed in shared resource pools. Consequently, efficient utilization of CPU resources is constrained by utilization of main memory resources or vice versa. This becomes pronounced when applications are memory or CPU intensive. In [22], Huawei proposed a composable high-throughput computing platform for big data applications in DCs. The platform disaggregates DC resources into independent pools of compute, memory, and I/O within each rack. Compute pools comprise of CPU cores and local memory i.e. partial disaggregation while the memory pool offers shared remote memory for resources across the rack. Software defined management plane and controllers in each pool enables centralized monitoring and orchestration of resources in each pool across the DC while the data plane controls data access operations. An optical network comprising of centralized top of rack (ToR) switch and optical links, was proposed as communication network fabric for the platform to support high throughput, low latency, low signal degradation and increased energy efficiency.

In the academic community, several authors [4]–[7], [9], [23]–[26] studied the disaggregation of CPU and memory resources into independent pools in addition to separation of IO and storage. Some of these authors [5], [7]–[9] studied the feasibility of complete disaggregation under a software defined management and configuration systems while others [4], [6], [24]–[27] proposed optical network topologies for disaggregated DCs. The authors in [5] formulated an integer linear programming model and developed simulated annealing based algorithm to optimally allocated virtual DC requests in a rack-scale DC using the optical DC interconnect proposed in [24]. The model compared the benefits of disaggregated DCs to traditional DCs using virtual machine (VM) rejection and resource utilization as the primary metrics. Other metrics, such as power consumption of DC network was not studied.

The work in [28] developed an energy efficient MILP model which optimally placed VMs in a pod-scale composable DC. In [4], the same authors proposed an electro-optical (hybrid) network for pod-scale composable DC and considered network power consumption in the MILP model and comparable heuristic was used to compare pod-scale composable DC and traditional DC infrastructures. The authors suggested a rack-clustering concept for pod-scale composable DC to ensure minimal inter-resource latency. However, the network topology proposed for pod-scale composable DC was designed to achieve resource disaggregation and did not evaluate the scalability of the proposed network for warehouse DC. In addition, the authors of [4], [28] only studied pod-scale composable DC which may be impracticable with state of art communication technologies according to [3]. On the other hand, the authors give less attention to disaggregation of DC resources at rack-scale which is more practical. The succeeding sections of this paper compare rack-scale and pod-scale approaches to resource disaggregation in composable DC under different network topologies.

In this paper, we extend our previous work in [29] where an initial study of the energy efficiency of networks for composable DC infrastructures was conducted. This paper extends our previous work in the following ways:

- A review of networks that are suited for composable DCs is made.
- A complete MILP model is given for the first time.
- Specific reference (electrical, optical and hybrid) network topologies are adopted to support the range of composable DCs considered. This contrasts with [29] where we did not consider any specific network topology.
- The impact of placing CPU and memory intensive workload types is studied.
- Adoption of micro-services to form integrated workloads in composable DCs is also considered in parallel with monolithic workloads.
- A heuristic that performs energy efficient placement of workloads in composable DCs is introduced for the first time.
- Finally, comprehensive discussions of results are made in this paper.

## III. REVIEW OF NETWORKS FOR COMPOSABLE DCS

Disaggregation of DC resources into independent resource pool to form composable infrastructure is only feasible when a network topology to support substantial growth in network traffic and near-conventional inter-resource access latency is available [6], [7], [30]-[31]. A few such network topologies have been proposed in existing literature for composable DC infrastructure adopting both rack-scale and pod-scale disaggregation. These network topologies adopt electrical, hybrid (electro-optic) and optical switching components and links. In this section, we review some of these network topologies and classify them based on the type of network components i.e. optical/electrical/hybrid adopted. Because the adoption of optical links is a standard practice in large-scale DCs, the use of optical links, as media to interconnect DC resource has no impact on our classification. Thus, hybrid network topologies must utilize both optical and electrical switches.

### A. Electrical Network Topologies

In [32], Intel proposed a multi-tiered Ethernet switching topology to interconnect partially disaggregated resources in the Intel Rack Scale Design (RSD) reference model. High versatility and low cost of Ethernet switches (due to their wide deployment) are the primary motivation given for their adoption in the network topology. As shown in Fig. 3, each homogenous resourced node in each rack of the reference model connects to a ToR Ethernet switch via a 10 GbE Network Interface Card (NIC) on the node. ToR switches in each rack of a pod connect to an aggregation (DC) switch that provides connectivity between intra-pod and inter-pod ToR switches. Aggregation switches also provide connectivity to external networks. Optionally, the Intel RSD supports a lower network tier where drawer Ethernet switches within the rack act as intermediate switches between resource nodes and the ToR switch within a rack. Intel RSD supports optical links at all tiers of the proposed network topology.

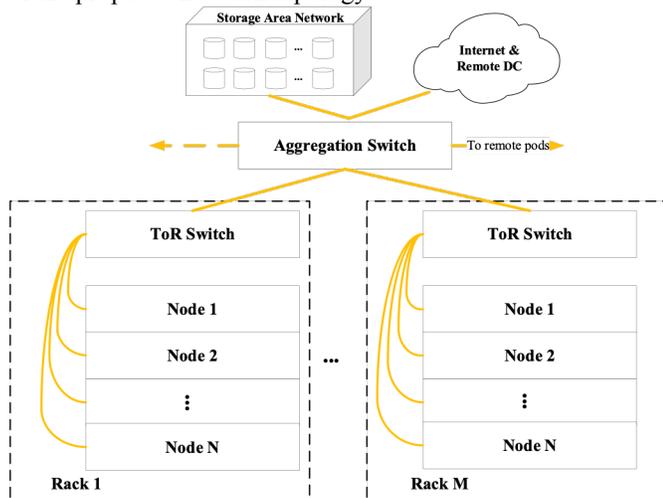

Fig. 3. An electrical network topology proposed for Intel RSD.

The authors in [22], also proposed a network topology for rack-scale composable DC infrastructures. This topology relies on electrical ToR switch (within each rack), which is the connection hub for all cloud controllers (present in each homogenous resourced node within the rack). Within the rack, the topology may also support a full-mesh or torus connectivity between cloud controllers on different resource nodes while a multi-tier switch-based network is proposed between ToR switches. In other literature [7], [33], the adoption of electrical switches such as PCIe and InfiniBand switches are proposed for DC infrastructures with composable I/O, storage, and accelerators.

However, the performance of latency sensitive applications running in DC infrastructures with separate CPU and memory components may significantly degrade due to the relatively high switch access latency of electrical switches. This is about 100ns for an InfiniBand switch [25], [30], 2-10μs for a leaf and spine switch [7] and approximately 150 ns [7] for a PCIe Gen3 switch. Recently, the Gen-Z consortium proposed high bandwidth sub-100ns latency electrical switches capable of interconnecting separated CPU and memory components over a switching fabric [34]. Such electrical switches can mitigate performance degradation of latency sensitivity applications. The Gen-Z consortium [35] is an industry led consortium with the objective of designing a new computer architecture that supports the disaggregation of CPU and memory components. Gen-Z supports the interconnection of disaggregated compute components via both direct and switch fabric attached approaches.

### B. Optical Network Topologies

To leverage on the advantages of optical network topologies over electrical network topologies such as data rate and modulation format agnosticism (i.e. high-speed transparent communication), increase of multiplexing domains and energy efficiency, some optical network topologies have been proposed for composable DC infrastructures. In these topologies, traffic between disaggregated compute resource components solely traverses optical components and links.

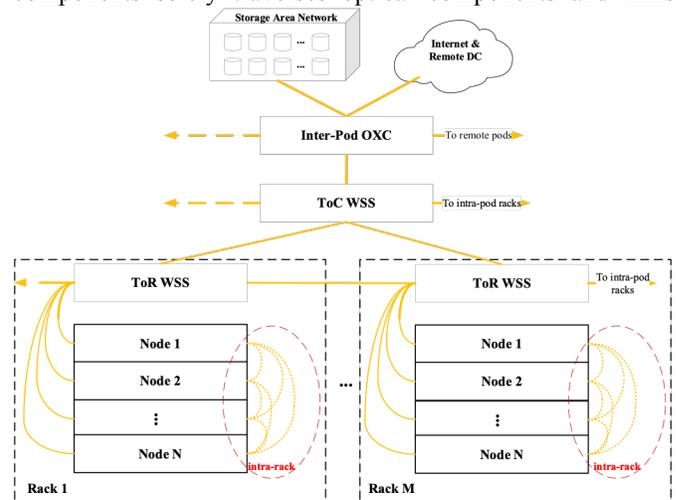

Fig. 4. An optical network topology for composable DCs - EVROS.

The work in [24]–[26] proposed optical network topologies for composable DC infrastructure at rack-scale. These topologies adopt active wavelength selective switches (WSSs) or passive AWGRs based switches and optical circuit switches (OCS) along inter-resource communication paths while bespoke FPGA-based switch interface cards (SICs) present on each homogeneous resource node are responsible for packet forwarding between nodes within the composable DC. In the first variant of the network topology called DORIS [26], WSS-based switches perform the function of a ToR switch by inter-connecting resource blades within the same racks and providing links to other WSS-based ToR switches in adjacent racks. Each ToR switch also connects to a higher tier WSS-based top of cluster (ToC) switch that in turns connects to optical fibre switches called inter-DC switches. DORIS supports direct and indirect (via the ToC switch) communication between intra-cluster racks while communication between clusters and to the internet is performed via optical inter-DC switches.

A second variation of the topology called EVROS [24], as shown in Fig. 4, introduced an intra-rack backplane which provides full mesh connectivity between intra-rack resource nodes via multi-port bespoke NICs also called SICs. EVROS adopts low propagation delay hollow-core photonic bandgap fibre (HC-PBGF) [36] for all links in the network topology. Authors in [25] enhanced the EVROS network topology via the use of configurable optical switches to replace ToC switches and inter-DC switches. These configurable switches have both OCS and optical packet switch (OPS) modules that can be configured on-demand to complement one another. Adoption of a remote SDN controller and SDN agents (embedded into SICs and architecture on demand (AoD) optical backplanes) in the DCN topology makes it SDN-enabled. The programmable AoD optical backplane acting as a ToC switch in each cluster is configurable on-demand to implement OCS or OPS connectivity between adjacent racks in the cluster, thus replacing the interconnection between ToR switches of adjacent racks.

A multi-tier optical network topology similar to the Intel RSD network topology was proposed for the rack-scale structure of the dRedBox project in [6], [37]. The rack-scale structure comprises of homogenous resource modules that are arranged into trays, which are placed in racks across the composable DC infrastructure. Each resource module possesses high-speed optical transceivers that provide an interface to the network topology. CPU resource modules are enhanced with FPGA programmable logic that performs layer 2 switching functionalities in the network topology and enable CPU modules to be used as intermediate switching nodes. The lower tier of the network topology is the intra-tray network (within each tray) that is formed by connections between intra-tray homogenous resource modules and one or more optical edge of tray (EOT) switches. Each EOT switch within a rack connects to a high radix optical ToR circuit switch to form the second tier of the network topology. Optical switches that reside above the second tier of the topology provide inter-rack/cluster communication within the DC. Further modifications to this DCN topology in [38] introduces a layer 1 electronic circuit switch within each tray that connects to all resource modules within the tray, while memory and accelerator modules are also embedded with FPGA programmable logic to enable them to participate in layer 2 switching functions.

### C. Hybrid Network Topologies

Other network topologies for composable DC infrastructures adopted both electrical and optical components; hence, they are hybrid network topologies. A hybrid network topology for rack-scale composable DC infrastructure was proposed in [30]. Intra-rack communication between compute and remote memory blades is achieved via a rack-local optical switch, which serves as a fast-optical backplane. A second (generic) backplane that begins in the rack and extends into a hybrid leaf-spine (with both electrical and optical spine switches) topology carries outward bound communications (such as CPU-IO, RAM-IO, CPU-disk, and RAM-disk traffic) of the composable DC infrastructure. Another hybrid network architecture was proposed for pod-scale composable DC infrastructure in [39]. This architecture deploys two tier switches in each homogenous resource rack as illustrated in Fig. 5. The first tier comprises of electrical switches, which directly connect to each intra-rack resource node via optical links, switches of the first-tier interface with ToR non-blocking optical switches, which provide full mesh connectivity between adjacent racks. Because layer 1 electronic circuit switches are adopted within each tray, the network topology proposed for dRedBox in [38] can also be classified as a hybrid network topology.

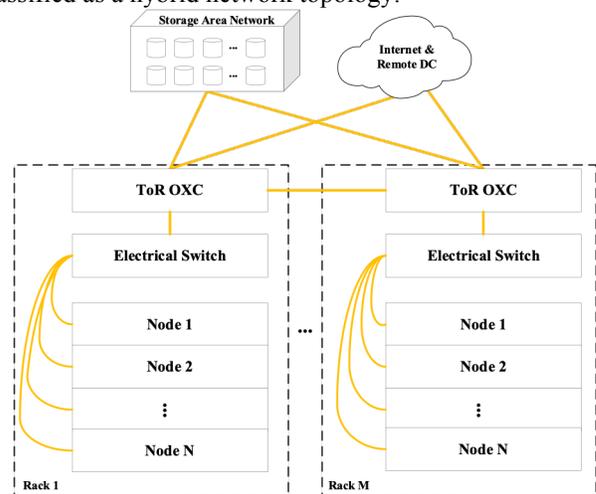

Fig. 5. A hybrid network topology for composable DC.

In this paper, our objective is to compare network topologies proposed for composable DCs under different scenarios. We select a representative topology each for electrical, optical and hybrid topologies discussed above. We adapt each representative topology to support traditional, rack-scale and pod-scale composable DCs as required. The multi-tier network topology of the Intel RSD is selected as the default electrical network topology. The EVROS network topology is adopted as the reference optical network topology, while the network topology proposed in [39] is the reference hybrid network topology.

## IV. ENERGY EFFICIENT WORKLOADS PLACEMENT

### A. Infrastructure Setup

The reference composable DC infrastructures are setup as follows; a DC comprise of one or more pods; each pod comprises of one or more racks and each rack comprises of one or more resource nodes. Under physical resource disaggregation, each rack in a pod-scale DC comprise of one type (i.e. CPU or memory) of resource only. Hence, a logical server can only be formed at the pod-level. On the other hand, under physical resource disaggregation, each node placed in a rack of rack-scale DC comprise of homogenous resources but each node in the rack can hold different resource types. Therefore, the lowest level for a logical server formation is at the rack-level. Therefore, a logical server can be formed at the rack-level. Each rack of a traditional DC comprises of heterogeneous nodes, each node holds CPU and memory resource components. Therefore, the lowest level for a logical server formation is at the node-level. Therefore, a logical server can be formed within a node in the rack.

It is important to note that a memory resource component is used in two different contexts in the infrastructure setup. In the first, a memory resource component performs the function of RAM as defined in conventional computer architecture. In another context, a memory resource component also performs the function of a storage device because in-memory computing is assumed for groups of workloads deployed in the DC. It is assumed that such groups of workloads perform in-memory data shuffle via remote direct memory access (RDMA) to reduce the impact such data exchange on the CPU and the operating system. Hence, inter-memory traffic is created in the network based on the placement of the memory resource demand of workloads in the DC.

### B. MILP Model Description

A mixed integer linear programming (MILP) model is developed to minimize the total power consumption of composable DC infrastructures which employ the reference electrical, hybrid or optical network topology. Given resource allocation in composable DC infrastructures and workload templates generated by an orchestrator sitting above the physical infrastructure layer, the model selects the optimum locations for each type of resource demanded by each workload so that the total DC power consumption is minimized.

In the modelled DC infrastructure, poor utilization of direct attached storage is mitigated by centralizing storage devices in remote systems such as a SAN and data is retrieved over a unified network topology that supports both local area network traffic and SAN traffic via a bespoke offload NIC. CPU resource components used in the model have sufficient local cache to support remote memory access after compute disaggregation. Additionally, un-capacitated network state is assumed for all reference network topology to ensure fair comparison. This also allows further simplification of the model as it can be assumed that network traffic is routed through the shortest path as expected in a best-case scenario for each reference network topology. The sets, parameters and variables of the MILP model are introduced as follows.

**Sets:**

| | |
|---|---|
| $N$ | Set of nodes of resources |
| $R$ | Set of racks in the DC |
| $P$ | Set of Pods in the DC |
| $CR$ | Set of CPU resources |
| $MR$ | Set of memory resources |
| $W$ | Set of workloads |
| $D$ | Set of traffic directions |

**DC Compute Parameters:**

| | |
|---|---|
| $C_j$ | Capacity of CPU module $j \in CR$ |
| $CPmax_j$ | Maximum power consumption of CPU module $j \in CR$ |
| IC | Idle power consumption as a fraction of maximum CPU power consumption |
| $\Delta C_j$ | Power factor of CPU module $j \in CR$; $\Delta C_j = \frac{CPmax_j - IC \cdot CPmax_j}{C_j}$ |
| $M_j$ | Capacity of memory module $j \in MR$ |
| $MPmax_j$ | Maximum power consumption of memory module $j \in MR$ |
| IM | Idle power as a fraction of maximum memory power |
| $\Delta M_j$ | Power factor of memory module $j \in MR$; $\Delta M_j = \frac{MPmax_j - IM \cdot MPmax_j}{M_j}$ |
| $CinN_{jn}$ | $CinN_{jn} = 1$, If CPU $j \in CR$ is placed in node $n \in N$. Otherwise $CinN_{jn} = 0$ |
| $MinN_{jn}$ | $MinN_{jn} = 1$ if RAM $j \in MR$ is placed in node $n \in N$, Otherwise $MinN_{jn} = 0$ |
| $NinR_{nr}$ | $NinR_{nr} = 1$, If node $n \in N$ is placed in rack $r \in R$, otherwise $NinR_{nr} = 0$ |
| $RinP_{rp}$ | $RinP_{rp} = 1$, If rack $r \in R$ is placed in pod $p \in P$, otherwise $RinP_{pr} = 0$ |
| $WC_w$ | CPU capacity required by workload $w \in W$ |
| $WM_w$ | Memory capacity required by workload $w \in W$ |
| $MaxLat_w$ | Maximum latency supported by DC infrastructure for workload $w \in W$ |
| $ICLat_{cm}$ | CPU-Mem latency between CPU component $c \in CR$ and memory component $m \in MR$. Inter-component latency. |
| $Q$ | A big number (100000) |
| $G$ | A big number (1000) |
| $\alpha$ | A weighing factor in Watts which specifies the cost per blocked workload |

**DC Network Parameters:**

| | |
|---|---|
| $OXCPC$ | Static power consumption of optical cross-connect (W) |
| $WSSPC$ | Static power consumption of WSS-based TOR switch (W) |
| $ESebp$ | Load proportionate energy of electrical switch (J/b) |
| $ESIPC$ | Idle power consumption of electrical switch (W) |
| $A$ | Number of aggregation switches. $A \geq 1$; $\frac{A}{NAR}$ is a fixed aggregation ratio. |
| $B$ | Number of inter-pod cross connects. $B \geq 1$; $\frac{B}{NAP}$ is a fixed ratio. |
| $TCM_{wx}$ | CPU-Memory (RAM) traffic (in b/s) of workload $w \in W$ in direction $x \in D$. |
| $TCI_{wx}$ | CPU-IO traffic of workload $w \in W$ in direction $x \in D$. |
| $TRI_{wx}$ | Memory (storage)-IO traffic of workload $w \in W$ in direction $x \in D$. |
| $IMT_{sd}$ | Inter-memory (storage) traffic between source workload $s \in W$ and destination workload $d \in W$. |

$ELECU_{sd}$ — Electrical network load proportional energy per bit (J/b) due to traffic from CPU component $s \in PR$ to memory (RAM) component $d \in MR$.

$ELECD_{sd}$ — Electrical network load proportional energy per bit (J/b) due to traffic from memory (RAM) component $s \in MR$ to CPU component $d \in PR$.

$ELEC_{sd}$ — Electrical network load proportional energy per bit (J/b) due to traffic between memory (storage) component $s \in MR$ and memory (storage) component $d \in MR$.

$ENSLEC$ — Electrical network load proportional energy per bit (J/b) due to north-south traffic.

$HLECU_{sd}$ — Hybrid network load proportional energy per bit (J/b) due to traffic from CPU component $s \in PR$ to memory (RAM) component $d \in MR$.

$HLECD_{sd}$ — Hybrid network load proportional energy per bit (J/b) due to traffic from memory (RAM) component $s \in MR$ to CPU component $d \in PR$.

$HLEC_{sd}$ — Hybrid network load proportional energy per bit (J/b) due to traffic between memory (storage) component $s \in MR$ and memory (storage) component $d \in MR$.

$HNSLEC$ — Hybrid network load proportional energy per bit (J/b) due to north-south traffic.

$OLECU_{sd}$ — Optical network load proportional energy per bit (J/b) due to traffic from CPU component $s \in PR$ to memory (RAM) component $d \in MR$.

$OLECD_{sd}$ — Optical network load proportional energy per bit (J/b) due to traffic from memory (RAM) component $s \in MR$ to CPU component $d \in PR$.

$OLEC_{sd}$ — Optical network load proportional energy per bit (J/b) due to traffic between memory (storage) component $s \in MR$ and memory (storage) component $d \in MR$.

$ONSLEC$ — Optical network load proportional energy per bit (J/b) due to north-south traffic.

**Variables:**

$WCL_{wj}$ — $WCL_{wj} = 1$ indicates that processing requirements of workload $w \in W$ are served by CPU $j \in CR$. Otherwise, $WCL_{wj} = 0$

$WML_{wj}$ — $WML_{wj} = 1$, indicates that memory (RAM) request of workload $w \in W$ is served by RAM $j \in MR$. Otherwise, $WML_{wj} = 0$

$CA_j$ — $CA_j = 1$, if CPU $j \in CR$ is active. Otherwise, $CA_j = 0$

$MA_j$ — $MA_j = 1$, if RAM $j \in MR$ is active. Otherwise, $MA_j = 0$

$PS_p$ — $PS_p = 1$, if pod $p \in P$ is active. Otherwise, $PS_p = 0$

$RS_r$ — $RS_r = 1$, if rack $r \in R$ is active. Otherwise, $RS_r = 0$

$S_w$ — Indicates the state of workload $w \in W$ i.e. served or unserved. $S_w = 1$, if workload $w$ is served. Otherwise, $S_w = 0$

$\beta_w$ — Indicates the state of workload $w \in W$ i.e. rejected or active. It is the opposite of $S_w$, $\beta_w = 1 - S_w$

$NAR$ — Number of active racks

$NAP$ — Number of active pods

$H_{wr}$ — $H_{wr} = 1$, if CPU resource demand of workload $w \in W$ is placed in rack $r \in R$. Otherwise, $H_{wr} = 0$

$F_{wr}$ — $F_{wr} = 1$, if memory resource demand of workload $w \in W$ is placed in rack $r \in R$. Otherwise, $F_{wr} = 0$

$A_{wp}$ — $A_{wp} = 1$, if CPU resource demand of workload $w \in W$ is placed in pod $p \in P$. Otherwise, $A_{wp} = 0$

$B_{wp}$ — $B_{wp} = 1$, if memory resource demand of workload $w \in W$ is placed in pod $p \in P$. Otherwise, $B_{wp} = 0$

$Y_{wcm}$ — Indicates the CPU-memory pair used to provision workload $w \in W$. $Y_{wcm} = 1$ if CPU $c \in CR$ and memory $m \in MR$ host CPU and memory resource demands of workload $w \in W$ respectively. Otherwise, $Y_{wcm} = 0$.

$\gamma_{sd}^{xy}$ — $\gamma_{sd}^{xy} = 1$ if memory resource demand of source workload $s \in W$ is placed in memory component $x \in MR$ and memory resource demand of destination workload $d \in W$ is placed in memory component $y \in MR$. Otherwise, $\gamma_{sdp} = 0$

The creation of the CPU requirements of workload $w$, $S_w$, can be related to the workload placement using:

$$S_w = \sum_{j \in CR} WCL_{wj} \tag{1}$$
$$\forall\, w \in W$$

The number of active racks is the DC can be derived from that state of each rack using:

$$NAR = \sum_{r \in R} RS_r \tag{2}$$

The number of active pods is the DC can be derived from that state of each pod using:

$$NAP = \sum_{p \in P} PS_p \tag{3}$$

The placement of the CPU resource requirement of workload $w$ into rack $r$ can be derived using:

$$H_{wr} = \sum_{n \in N} \sum_{j \in CR} WCL_{wj} CinN_{jn} NinR_{nr} \tag{4}$$
$$\forall\, w \in W, \forall\, r \in R$$

The placement of the memory resource requirement of workload $w$ into rack $r$ can be derived using:

$$F_{wr} = \sum_{n \in N} \sum_{j \in MR} WML_{wj} MinN_{jn} NinR_{nr}$$
$$\forall w \in W, \forall r \in R$$
(5)

The placement of the CPU resource requirement of workload $w$ into pod $p$ can be derived using:

$$A_{wp} = \sum_{r \in R} \sum_{n \in N} \sum_{j \in CR} WCL_{wj} CinN_{jn} NinR_{nr} RinP_{rp}$$
$$\forall w \in W, \forall p \in P$$
(6)

The placement of the memory resource requirement of workload $w$ into pod $p$ can be derived using:

$$B_{wp} = \sum_{r \in R} \sum_{n \in N} \sum_{j \in MR} WML_{wj} MinN_{jn} NinR_{nr} RinP_{rp}$$
$$\forall w \in W, \forall p \in P$$
(7)

Given each network topology, the load proportional power consumption of network components traversed and some variables, the power consumption of electrical, hybrid and optical network topologies are derived as follows:

**Electrical network Topology**

$$TNPC = \sum_{c \in CR} \sum_{m \in MR} \sum_{w \in W} Y_{wcm}(TCM_{w1}ELECU_{cm} + TCM_{w2}ELECD_{cm})$$
$$+ \sum_{x \in MR} \sum_{y \in MR: x \neq y} \sum_{s \in W} \sum_{d \in W:s \neq d} \gamma_{sd}^{xy} IMT_{sd}ELPC_{xy}$$
$$+ ENSLEC \left( \sum_{w \in W} \sum_{c \in CR} WCL_{wc}(TCI_{w1} + TCI_{w2}) \right.$$
$$\left. + \sum_{w \in W} \sum_{m \in MR} WML_{wm}(TRI_{w1} + TRI_{w2}) \right)$$
$$+ (NAR + A)ESIPC$$
(8)

**Hybrid Network Topology**

$$TNPC = \sum_{c \in CR} \sum_{m \in MR} \sum_{w \in W} Y_{wcm}(TCM_{w1}ELECU_{cm} + TCM_{w2}ELECD_{cm})$$
$$+ \sum_{x \in MR} \sum_{y \in MR: x \neq y} \sum_{s \in W} \sum_{d \in W:s \neq d} \gamma_{sd}^{xy} IMT_{sd}ELPC_{xy}$$
$$+ ENSLEC \left( \sum_{w \in W} \sum_{c \in CR} WCL_{wc}(TCI_{w1} + TCI_{w2}) \right.$$
$$\left. + \sum_{w \in W} \sum_{m \in MR} WML_{wm}(TRI_{w1} + TRI_{w2}) \right)$$
$$+ (NAR + A)ESIPC$$
(9)

**Optical Network Topology**

$$TNPC = \sum_{c \in CR} \sum_{m \in MR} \sum_{w \in W} Y_{wcm}(TCM_{w1}OLECU_{cm} + TCM_{w2}OLECD_{cm})$$
$$+ \sum_{x \in MR} \sum_{y \in MR: x \neq y} \sum_{s \in W} \sum_{d \in W:s \neq d} \gamma_{sd}^{xy} IMT_{sd}OLEC_{xy}$$
$$+ ONSLEC \left( \sum_{w \in W} \sum_{c \in CR} WCL_{wc}(TCI_{w1} + TCI_{w2}) \right.$$
$$\left. + \sum_{w \in W} \sum_{m \in MR} WML_{wm}(TRI_{w1} + TRI_{w2}) \right)$$
$$+ NARWSSPC + OXCPC(NAP + B)$$
(10)

**Total CPU Power Consumption**

Total power consumption of CPU resources in the composable DC (TCPC) is derived as follows.

$$TCPC = \sum_{j \in CR} \sum_{w \in W} \left( (ICCPmax_j CA_j) + (\Delta C_j WCL_{wj} WC_w) \right)$$
(11)

**Total Memory Power consumption**

Total power consumption of memory resources in the composable DC (TMPC) is derived as follows.

$$TMPC = \sum_{j \in MR} \sum_{w \in W} \left( (IMMPmax_j MA_j) + (\Delta M_j WML_{wj} WM_w) \right)$$
(12)

The MILP model's objective function minimizes the total power consumption of CPU resources, memory resources, and the network topology used across the DCs and the number of rejected workloads in scenarios where some workloads cannot be provisioned. $\alpha$ is the cost (measured in Watts) associated each rejected workload.

**Minimize:**

$$TCPC + TMPC + TNPC + \alpha \left( \sum_{w \in W} \beta_w \right)$$
(13)

**Subject to the following constraints:**

$$\sum_{w \in W} WC_w WCL_{wj} \leq C_j$$
$$\forall j \in CR$$
(14)

$$\sum_{w \in W} WM_w WML_{wj} \leq M_j$$
$$\forall j \in MR$$
(15)

Constraints (14) and (15) denote resource capacity constraints for each unit of CPU and memory component in the DC.

$$\sum_{j \in CR} WCL_{wj} \leq 1$$
$$\forall w \in W$$
(16)

$$\sum_{j \in MR} WML_{wj} \leq 1$$
$$\forall w \in W$$
(17)

Constraints (16) and (17) limit the maximum number of components that can host CPU and memory resource requests of a workload to one, this is because neither replication nor slicing of workloads is permitted. These constraints also allow workloads to be rejected in scenarios where resource capacity is limited.

$$\sum_{j \in CR} WCL_{wj} = \sum_{j \in MR} WML_{wj} \qquad (18)$$
$$\forall w \in W$$

Constraint (18) ensures that an active workload's CPU and memory resource requirements are satisfied. Otherwise the workload is inactive.

$$G \sum_{w \in W} WCL_{wj} \geq CA_j \qquad (19)$$
$$\forall j \in CR$$

$$\sum_{w \in W} WCL_{wj} \leq QCA_j \qquad (20)$$
$$\forall j \in CR$$

$$G \sum_{w \in W} WML_{wj} \geq MA_j \qquad (21)$$
$$\forall j \in MR$$

$$\sum_{w \in W} WML_{wj} \leq QMA_j \qquad (22)$$
$$\forall j \in MR$$

Constraints (19) - (22) determine each CPU and memory resource state, this depends on utilization of the resource to satisfy resource requirements of served workloads.

$$\sum_{w \in W} H_{wr} + F_{wr} \geq RS_r \qquad (23)$$
$$\forall r \in R$$

$$\sum_{w \in W} H_{wr} + F_{wr} \leq QRS_r \qquad (24)$$
$$\forall r \in R$$

Constraints (4.23) and (4.24) determine the state of each rack, this depends on the utilization of CPU or memory resource in the rack to satisfy resource requirements of served workloads.

$$G \sum_{w \in W} A_{wp} + B_{wp} \geq PS_p \qquad (25)$$
$$\forall p \in P$$

$$\sum_{w \in W} A_{wp} + B_{wp} \leq QPS_p \qquad (26)$$
$$\forall p \in P$$

Constraints (25) and (26) determine the state of each pod, this depends on the utilization of CPU or memory resource in the pod to satisfy resource requirements of served workloads.

$$Y_{wcm} = WCL_{wc} WML_{wm} \qquad (27)$$
$$\forall w \in W, \forall c \in CR, \forall m \in MR$$

Constraint (27) gives the relationship between components hosting processing and memory resource demands of a given workload i.e. $Y_{wcm}$ which is a product of $WCL_{wc}$ and $WML_{wm}$.

$$Y_{wcm} \leq WCL_{wc} \qquad (28)$$
$$\forall w \in W, \forall c \in CR, \forall m \in MR$$

$$Y_{wcm} \leq WML_{wm} \qquad (29)$$
$$\forall w \in W, \forall c \in CR, \forall m \in MR$$

$$Y_{wcm} \geq WCL_{wc} + WML_{wm} - 1 \qquad (30)$$
$$\forall w \in W, \forall c \in CR, \forall m \in MR$$

Constraints (28) - (30) are used to linearize constraint (27), which is a product of two binary variables.

$$\sum_{c \in CR} \sum_{m \in MR} ICLat_{cm} Y_{wcm} \leq MaxLat_w \qquad (31)$$
$$\forall w \in W$$

Constraint (31) ensures that the inter-component latency between the CPU and memory components hosting a workload's resource demands does not exceed the set maximum CPU-Memory latency for a given type of composable DC infrastructure. This constraint is combined with the allocation of components to control the type of composable DC infrastructure under consideration and enforces resource locality.

$$\gamma_{sd}^{xy} = WML_{sx} WML_{dy} \qquad (32)$$
$$\forall s \in W, \forall d \in W, \forall x \in MR, y, \forall \in MR$$

$$\gamma_{sd}^{xy} \leq WML_{sx} \qquad (33)$$
$$\forall s \in W, \forall d \in W, \forall x \in MR, y, \forall \in MR$$

$$\gamma_{sd}^{xy} \leq WML_{dy} \qquad (34)$$
$$\forall s \in W, \forall d \in W, \forall x \in MR, y, \forall \in MR$$

$$\gamma_{sd}^{xy} \geq WML_{sx} + WML_{dy} - 1 \qquad (35)$$
$$\forall s \in W, \forall d \in W, \forall x \in MR, y, \forall \in MR$$

Constraint (32) gives the relationship between two memory components ($x \in MR$ and $y \in MR$) hosting memory resource demands of workloads ($s \in W$ and $d \in W$) respectively i.e. $\gamma_{sd}^{xy}$ which is a product of binary variables $WML_{sx}$ and $WML_{dy}$. Constraints (33) - (35) are used to linearize constraint (32), which is a product of two binary variables.

## V. PERFORMANCE EVALUATION

The MILP model is used to study the performance of composable DCs that employ physical resource disaggregation at rack-scale and pod-scale relative to the performance of traditional DC infrastructure. Heterogeneous CPU and memory resource component are adopted to reflect heterogeneity of resources in production DCs. Three classes of servers illustrated in Table I are adopted to form a range of heterogeneous resourced DCs. To minimize the execution time of the MILP model which grows as the size and complexity of the problem increases, small DCs comprising of 24 servers (i.e. 24 CPU and 24 memory components) are considered; 8 servers from each server class are selected. In the traditional DC, servers maintain their single-box architecture while in rack-scale and pod-scale DCs, components of each server are physically disaggregated accordingly.

TABLE I
SERVER CLASSIFICATION

| Server Class | CPU capacity (Peak power) | Memory capacity (Peak power) |
|---|---|---|
| High Performance | 3.6 GHz (130 W) [28] | 32GB (40W) [28] |
| Standard | 2.66 GHz (95 W) [28] | 24 GB (30.72 W) [28] |
| Legacy | 2.4 GHz (80 W) [28] | 8 GB (10.24 W) [28] |

Resources are allocated as follows:

- Each DC has 2 pods, a pod comprises of two heterogeneous/homogenous resourced racks, and each rack holds multiple homogenous or heterogeneous resourced nodes.
- Each DC comprise of 24 CPU and 24 memory components which are allocated to nodes within each rack.
- Each rack of the traditional DC is a heterogeneous rack that holds 6 heterogeneous nodes i.e. traditional servers (2 servers from each class of server defined).
- Each rack of rack-scale DC holds 3 homogenous nodes of CPU resources and 3 homogenous nodes of memory resources. Hence, rack-scale DC comprise of heterogeneous resourced racks, 2 of these heterogeneous racks are allocated to each pod.
- In a pod-scale DC, racks hold 6 homogenous nodes of CPU or memory resources i.e. each rack comprise of homogenous nodes of CPU or memory resources. Each pod comprises of 1 homogenous rack of CPU resources and 1 homogenous rack of memory resources.

Allocation of resource components in traditional, rack-scale, and pod-scale DCs determines the inter-component latency between any CPU and memory resource component pair in the DC. Table II gives the abstracted inter-component latency between pair DC components (i.e. CPU and memory) in traditional DCs and in a DC that is physically disaggregated DCs at rack-scale or pod-scale. Resource component allocation and the choice of maximum latency($MaxLat_w$) for each workload in Constraint (31) collectively determine the composable DC infrastructure being evaluated by the MILP model. For simplicity, it is assumed that each traditional server has a single CPU component and a single memory component before disaggregation. Hence, both CPU and memory components of a given server can share a common index. Therefore, $ICLat_{cm}$ is derived from resource component allocation i.e. $CinN_{jn}, MinN_{jn}, NinR_{nr}$ and $RinP_{rp}$ as described earlier. This derivation is guided by the latency abstraction in Table II. Similarly, allocation of resource components in DCs also aids the pre-computation of load proportional power consumption of network components traversed on each reference network topology considered i.e. $ELPCU_{sd}, ELPCD_{sd}, ELPC_{sd}, ENSLPC, HLPCU_{sd}, HLPCD_{sd}, HLPC_{sd}, HNSLPC, OLPCU_{sd}, OLPCD_{sd}, OLPC_{sd},$ and $ONSLPC$.

In addition to the fixed idle power consumption, each CPU and memory resource component has a linear load proportionate power profile. The "power factor" which represents the slope of the linear power profile is the active power consumed per resource capacity. This is the basis for calculating the load proportionate power of each active CPU and memory resource components. However, the power factor of a resource component alone does not give a full picture of its energy efficiency. This is because a resource component with low power factor may have little capacity; hence, it can only support a small volume of resource demand. Normalizing the power factor of components by their corresponding capacity gives a better measure of energy efficiency as shown in Fig. 6. Therefore, at full (100%) CPU resource utilization, high performance server (3.6 GHz) CPU is the most energy efficient, followed by standard server (2.66 GHz) CPU. Thereafter, the legacy server (2.4 GHz) CPU closely follow standard server CPU as shown in Fig. 6. Energy efficiency of fully utilized memory components also follow the same order. The idle power consumption of the CPU and memory resources is 70% of the maximum power. Overall, from Table I, CPU components peak power consumption is relatively higher than that of memory components. The maximum power and idle power as a fraction of maximum power of both CPU and memory components are given in Table III. Table III also gives the load proportionate energy per bit values of next-generation network interfaces (using silicon photonics technologies) at different tiers of DC infrastructures.

TABLE II
COMPOSABLE DC INTER-COMPONENT LATENCY ABSTRACTION

| DC type | Relative location of pair component | Abstraction values |
|---|---|---|
| Traditional | Pair components are in the same server/node | 1 |
| Rack-scale | Pair components are in different server within the same rack | 2 |
| Pod-scale | Pair components are in different racks within the same pod | 3 |
| DC-scale | Pair components are in different pods within the same DC | 4 |

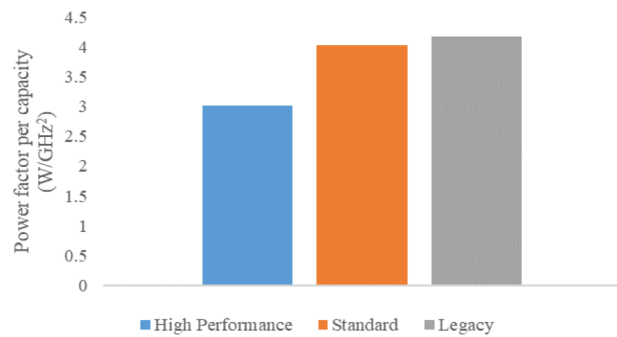

Fig. 6. Power factor per capacity of server CPU components

The power consumption of electrical and optical components in different candidate network topologies are evaluated using the parameters given in Table III. Because we have adopted a small DC infrastructure setup, parameters $A$ and $B$ which represents the number of aggregation switches in each pod and the number of inter-pod cross connects (in corresponding network topologies) respectively are both set to one. The load proportionate energy (28.28 pJ/b) of electrical switches is the ratio of dynamic power range (181W) of the switch to the maximum switching capacity (6.4 Tbps) of a Cisco Nexus 3132C-Z [40] electrical switch. The dynamic power range of a component given by the difference of the maximum and idle power consumption of that component. Programmable Finisar 4x16 WSS [41] (with maximum power consumption of 50W) is adopted as the WSS-based ToR switch in the optical network topology. An all-optical circuit switch [42] with maximum power consumption of 75W is deployed as the OXC and AOD switch in both optical and hybrid network topologies. These values and the energy efficiency values of next generation silicon photonics networks (adopted from [43]) are used as default parameters to evaluate the power consumption at the interface of resource components and nodes. Next-generation networks are expected to be more energy and cost efficient than today's networks; hence, the low values for next-generations networks as given in **Error! Reference source not found.**Table III. Relative to communication between nodes in the same or different racks over longer distances, on-board communication between components in compute node requires lower power consumption as a simple design is needed. Compared to long-distance communication between distributed DCs over a DC interconnect, inter-node communication between nodes within the DC is more energy efficient as shorter distances are travelled within the DC. Hence, less complex interfaces with lower power consumption are required for intra-DC networks relative to inter-DC networks.

TABLE III
DC COMPONENTS AND INTERFACES POWER CONSUMPTION

| Description | Value |
|---|---|
| Idle power as a fraction of maximum CPU power | 70% |
| Idle power as a fraction of maximum memory power | 70% |
| On-board network interface energy per bit (J/b) | 0.5 pJ/b [43] |
| Inter-rack network interface energy per bit (J/b) | 1 pJ/b [43] |
| Rack backplane network interface energy per bit (J/b) | 1 pJ/b [43] |
| Inter-DC network interface energy per bit (J/b) | 10 pJ/b [43] |
| Peak power consumption of optical circuit switch | 75 W [42] |
| Peak power consumption of WSS-based optical switch | 50 W [41] |
| Peak power consumption of electrical switch | 493 W [34] |
| Typical operating (idle) power of electrical switch | 312 W [34] |
| Load proportional energy of electrical switches | 0.028 pJ/b |

Heterogeneity of DC workloads is considered by adopting two classes of workloads i.e. CPU intensive and memory intensive workloads. Uniform distribution of resources intensity over specified ranges in Table IV is used to generate 5-20 monolithic workloads for each workload class. These ranges are selected relative to the maximum capacity of each resource type. It is assumed that cache coherent traffic is limited to each CPU and does not traverse the DC network fabric [11]. Except for inter-memory traffic resulting from in-memory data shuffle between workloads i.e. $IMT_{sd}$, fixed inter-resource communication traffics for each workload is considered (as given in Table IV) to ensure fair comparison between different composable DC infrastructures. Workloads are clustered into groups of five to represent groups of associated applications in conventional DCs such as worker and master nodes. Each workload group of associated workloads has one-to-one, one-to-many, many-to-many or mixed inter-memory traffic patterns between the workloads in that group. In most situation, it is assumed that inter-memory traffic is bandwidth intensive or non-bandwidth intensive over the range given in Table IV.

The described MILP model evaluates the impact of electrical, hybrid and optical network topologies on the performance of rack-scale or pod-scale composable DCs relative to traditional DC infrastructure. The MILP model is solved using the 64-bit AMPL/CPLEX solver on the ARC3 supercomputing node with 24 CPU cores and 128 GB of memory [44]. The MILP model results analysis consider metrics such as CPU, memory and network power consumption, number of active DC resources and average active resource utilization. Average active DC resource utilization represents the average utilization of all active CPU or memory resource components in the DC. The average utilization of network components such as switches is not considered. To obtain optimal results, the MILP model bin-packs workloads resource demands onto DC resources to achieve optimal resource power and utilization efficiencies within capacity and resource locality constraints while each workload's inter-resource communication (between CPU-memory component pairs for a workload) and inter-memory data shuffle traffic traverse different network topologies.

TABLE IV
WORKLOADS RESOURCE AND TRAFFIC DEMAND INTENSITY

| Resource Demand | CPU Intensive | Memory Intensive |
|---|---|---|
| CPU Demand (GHz) | 1 - 3 | 0.5 - 2 |
| Memory Demand (GB) | 4 - 8 | 6 - 24 |
| Uplink/Downlink CPU-Memory traffic (Gbps) | 120/100 | |
| Uplink/Downlink CPU-IO traffic (Gbps) | 2/1 | |
| Uplink/Downlink Memory-IO traffic (Gbps) | 2/1 | |
| Memory-memory traffic (Gbps) | Non-intensive 0-10Gbps or intensive 10-70Gbps | |

### A. CPU Intensive Workloads

Under all network topologies, the traditional DC infrastructure has the highest quantity of active DC resources (CPU and memory i.e. server) and the lowest average active memory resources utilization relative to other DC infrastructures considered as shown by results obtained when 20 CPU intensive workload are optimally provisioned. These observations are consistent with the widely reported challenges of provisioning monolithic workloads in traditional DCs which are characterized by disproportionate utilization of DC resource [2], [3], [5]. The CPU intensive nature of input workloads is responsible for high average active CPU resource utilization observed in traditional DC as shown in Fig. 7a**Error! Reference source not found.Error! Reference source not found.**.

For each number of CPU intensive workloads provisioned, the total CPU power consumption (TCPC) observed in traditional DC when the electrical network topology is deployed is equal to the TCPC observed when the hybrid network topology is deployed in the traditional DC as seen in Fig. 8a and 8b**Error! Reference source not found.**. The total memory power consumption (TMPC) also follows the same trend under both electrical and hybrid network topologies. On the other hand, the contributions of TCPC and/or TMPC to the total DC power consumption (TDPC) under the optical network topology is occasionally to some extent different from those observed under the electrical or hybrid network topology. Lower network power consumption per active rack in the optical network topology promotes the use of a different mix of active servers (CPU and memory resource components as seen in Fig. 7b and 7c) that are distributed across racks in the DC is responsible for lower TCPC and TMPC observed under the optical network topology.

In contrast, the presence of electrical switches (which have significant idle power consumption) in the lowest tiers of electrical and hybrid network topologies implies that there is a preference for consolidation of workloads into a few active racks over the use of servers distributed across racks in the DC for better energy efficiency. This leads to activation of few racks and ensures lower TDPC in traditional DCs. For example, when 20 CPU intensive workloads are optimally provisioned in traditional DC, the TCPC and TMPC of traditional DC with optical network topology is 6% and 11% respectively lower than the values reported for traditional DC with electrical or hybrid network topologies. Fig. 7b and 7c show the disparity in server (CPU and memory) resource usage under different network topologies when 20 CPU intensive workloads are provisioned.

The TNPC in traditional DC depends on the placement of workload resource demand into resource components. This is because workload resource demand placement determines the number of active racks and traffic that flows in the tiers of DCN topologies. As expected, TNPC changes with the network topologies as seen in Fig. 8 due to the variance in the power consumption profiles of components and interfaces that are present in the tiers of each network topology. In traditional DCs, high bandwidth inter-component traffics are node-limited i.e. they are restricted to on-board backplane of servers while inbound and outbound traffics of the DC (traffic between DC resources and remote systems) flow through higher tiers of each network topology. The contributions of node-limited traffic to the TNPC is small and constant across all network topologies in the traditional DCs due to relatively higher energy efficiency of on-board backplane compared to other tiers of each DCN topology as illustrated in Table III. Although lower bandwidth traffic traverses the electrical network topology in a traditional DC, the electrical network topology has the highest TNPC because of significant idle power consumption of electrical switches of the multi-tier topology. The TNPC of the hybrid network topology closely follows that of the electrical network topology because each rack in the DC also has an electrical switch with significant idle power consumption in addition to the fixed low power consumption of OXCs present in the higher tier of the hybrid topology. As shown in Fig. 8c, the optical network topology has the lowest TNPC because of fixed-low power consumption of poorly utilized high-capacity optical switches in the topology.

For varying number of CPU intensive workloads, the total DC power consumption in the traditional DC with electrical or hybrid network topology is largely subjective to TNPC, because TCPC and TMPC are relatively constant. Hence, a traditional DC that deploys an electrical network topology has the highest TDPC while the TDPC of traditional DCs with hybrid network topology follows. The TDPC in the traditional DC with optical network topology is the lowest because the adoption of optical switches in the topology ensure lower TNPC which in-turn encourages lower TCPC and TMPC as observed when 20 CPU intensive workloads are provisioned. Under varying number of CPU intensive input workloads considered, 10% average percentage reduction in TDPC was obtained when the hybrid network topology replaced the electrical network topology. Similarly, 27% average percentage reduction in TDPC was obtained when the optical network topology replaced the hybrid network topology under varying number of CPU intensive input workloads considered.

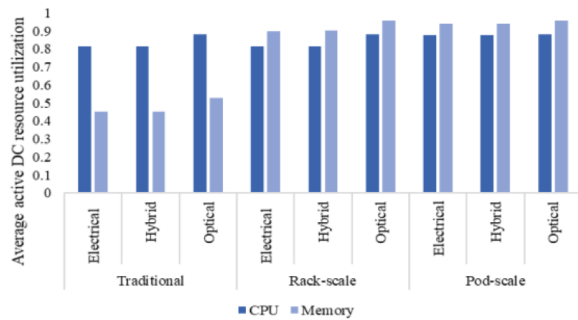

(a) Average active DC resource utilization

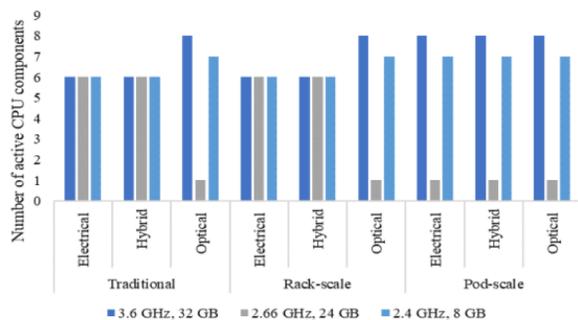

(b) Active CPU under 20 CPU intensive workloads

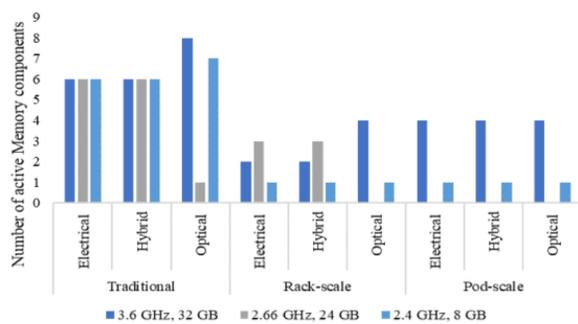

(c) Active memory under 20 CPU intensive workloads

Fig. 7. Average utilization and number of active DC resource components under 20 CPU intensive workloads.

Note that in the traditional DC, strict resource locality (required between CPU and memory components that host each workload) and resource capacity constraints collectively limits clustering of memory resource demands of workloads (that belong to the same workload group) into the same memory component. Hence, memory data shuffle traffic between memory components in different nodes has little impact on the placement of workloads in traditional DC when energy minimization is the goal. Memory data shuffle traffic exists between nodes if two workloads within the same workload group with inter-memory traffic ($IMT_{sd}$) are placed in different nodes.

Under varying number of CPU intensive workloads considered, physical disaggregation of traditional servers at rack-scale improves efficient usage of memory resources relative to the traditional DC via proportional usage of memory resource under all network topologies. Proportional usage of memory resources enable the reductions in the number and diversity of active memory resource components as shown in Fig. 7c and their corresponding TMPC as shown in Fig. 8**Error! Reference source not found.**. Both number of active memory resource components and their corresponding TMPC reduced by more than 50% when traditional DC servers are physically disaggregated at rack-scale under all network topologies considered. Relaxed inter-resource locality constraint in rack-scale DCs which led to improvements in average active memory resource utilization and energy efficient (i.e. proportional) usage of active memory resources in the DC is responsible for this. Hence, 32 GB and 24 GB memory components are activated when they can be highly utilized. Otherwise, 8GB memory components are activated if memory capacity constraints permit.

The CPU intensive nature of input workloads implies that there are limited opportunities to improve overall CPU power efficiency in the DC via physical disaggregation. For example, when 20 CPU intensive workloads were optimally provisioned in the DC with electrical, hybrid or optical network topology, the results show that there is no improvement in TCPC of rack-scale DC relative to the value obtained when a traditional DC with a similar network topology was used. As reported under traditional DC, the type of network topology adopted also determines optimum placement of workload resource demands in rack-scale DCs after physical disaggregation. Hence, relative to the electrical or hybrid network topologies in a rack-scale DC, revised placement of workload resource demands leads to lower TCPC (6% fall) and (further decrease in) TMPC (5%) when the optical network topology is used with minimal increase in TNPC. However, this revised placement of CPU and memory resource demands which delivers better energy efficiency may increase inter-rack and inter-pod network traffic because of additional inter-memory data shuffle traffic between memory components of workloads that belong to the same workload group, which are placed in different racks (or pods). In summary, in rack-scale DC where resource locality is limited to the rack, deployment of optical network topology achieves lower TMPC at the cost of higher network traffic and marginal increase in TNPC. In contrast, when the hybrid or electrical network topology is deployed, lower network traffic and TNPC is preferred at the cost of higher TMPC.

In the rack-scale DC, high bandwidth traffic between CPU and memory components is rack-limited, while low bandwidth DC north-south communication traverses higher tiers of the network topology adopted. Hence, as observed in the traditional DC, the electrical network topology has the highest TNPC in rack-scale DC while the TNPC of hybrid and optical network topologies follow in descending order as shown in Fig. 8. Because of relaxed inter-resource locality constraint between

CPU and memory components and the low memory resource demand that characterizes CPU intensive workloads, there is an increase in the number of instances in which memory resource demands of workloads (that belong to the same workload group) are collocated within the same memory component or the same homogenous memory node. This enables marginal reductions in TNPC of electrical, hybrid and optical network topologies alike under the rack-scale DC.

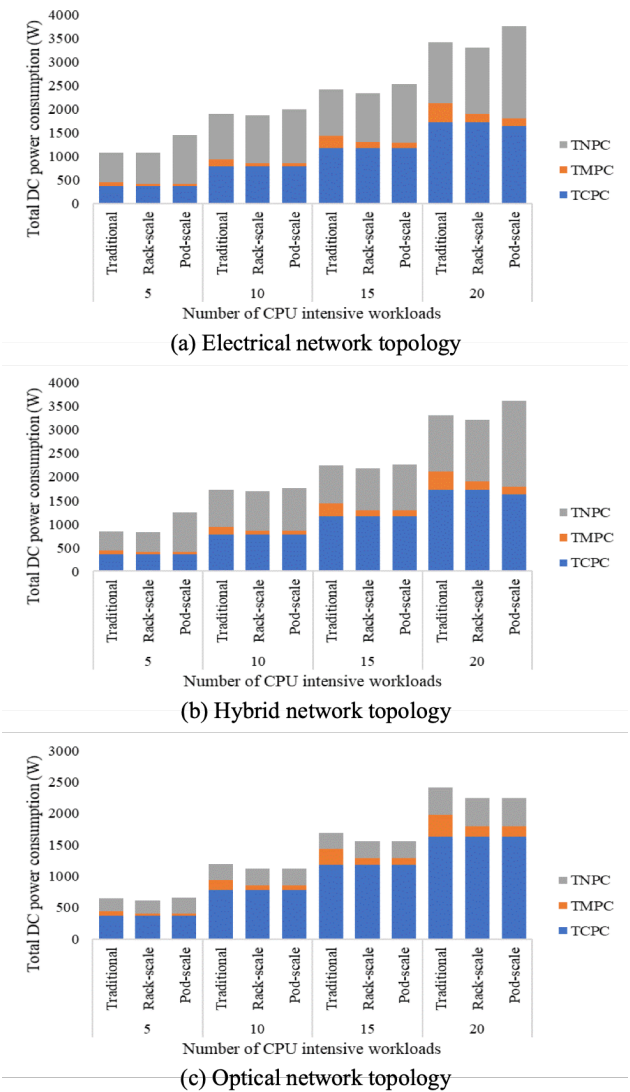

(a) Electrical network topology

(b) Hybrid network topology

(c) Optical network topology

Fig. 8. Total DC power consumption of composable DCs under CPU intensive workloads and different network topologies.

In the pod-scale DC, high-bandwidth traffic between CPU and memory components is pod-limited while low bandwidth DC north-south communication traverses all tier of the network topology in the DC. Hence, power consumption of electrical and hybrid network topologies in pod-scale DC increases significantly relative to traditional and rack-scale DCs as shown in Fig. 8. Sole adoption of power-hungry electrical switches in the electrical network topology resulted in very high TNPC relative to hybrid and optical network topologies as shown in Fig. 8. Unlike observation under traditional and rack-scale DCs, network topology does not inhibit optimal selection of CPU and memory resources in pod-scale DC. This is because all CPU-memory traffic must traverse inter-rack fabric due to the use of homogenous resourced racks in the pod-scale DC. As observed in the rack-scale DC, network power consumption resulting from inter-workload memory data shuffle is also significantly limited in pod-scale DC. This is achieved by placing memory resource demand of workloads of the same workload group in the same component or node. However, clustering of workloads memory resource demands into memory components or nodes also depends on memory capacity constraint, the power consumption of memory components and their corresponding impact on the total DC power consumption.

Comparison of TCPC and TMPC in rack-scale and pod-scale DCs shows that rack-scale DC can achieve similar performance as pod-scale DC as shown in Fig. 8. This is because the number and diversity of each resource type in each rack of rack-scale DC may be sufficient (e.g. when 5 or 10 CPU intensive workloads are optimally provisioned) to enable optimal benefits of resource disaggregation. However, as observed when 20 CPU intensive workloads are optimally provisioned in pod-scale DC with hybrid or electrical network topology, greater diversity and higher number of CPU resource components in homogenous racks of pod-scale DC relative to the situation in heterogeneous racks of rack-scale DC can enable better overall CPU power efficiency in pod-scale DC relative to the rack-scale DC. Hence, if the number and diversity of each resource type required by workloads is guaranteed during resource component allocation in rack-scale DC, similar TCPC and TMPC can be achieved under rack-scale and pod-scale DCs. In DCs that deploy electrical or hybrid network topology, satisfaction of the outlined requirement ensures lower TNPC in rack-scale DCs relative to the TNPC of pod-scale DCs. On the other hand, if an optical network topology such as EVROS is deployed, rack-scale DC's TCPC, TMPC and TNPC can match those of the pod-scale DC as seen in Fig. 8c when 10 CPU intensive workloads are provisioned in both rack-scale and pod-scale DCs that deploy the optical network topology and have equal number of active racks.

Generally, under varying number of CPU intensive workloads, the highest (7%) average percentage reduction in TDPC is achieved when rack-scale DC with optical network topology is adopted to replace the traditional DC with optical network topology. Compared to the traditional DC with hybrid network topology, 3% average percentage reduction in TDPC is achieved when rack-scale DC with hybrid network topology is adopted to replace a traditional DC with the same network topology. Compared to the traditional DC with electrical network topology, the average percentage reduction in the TDPC of rack-scale DC with similar network topology is 2%. It is also observed that only the optical network topology enabled reductions in the TDPC of traditional DC when the pod-scale DC (except when 5 CPU intensive workloads are optimally provisioned). With respect to reduction in TDPC, the results also show that physical disaggregation should be limited to the rack–scale to ensure TDPC reductions if electrical or hybrid network topology are to be deployed for disaggregated DC. Going further by disaggregating at pod-scale leads to significant (up to 50%) increase in the TNPC of electrical and hybrid network topologies which surpasses any savings in TCPC and TMPC derived from physical disaggregation of traditional DC at pod-scale.

## B. Memory Intensive Workloads

Results under memory intensive workloads further highlights trends identified when CPU intensive workloads were provisioned. Proportional usage of CPU and memory resources in disaggregated DCs (as seen in Fig. 9a) led to lower TCPC and TMPC as seen in Fig. 10. For instance, when 5 input workloads are provisioned across traditional, rack-scale, and pod-scale DCs, physical disaggregation of resources at rack-scale and pod-scale leads to reductions in TCPC and TMPC relative to the traditional DC. This is achieved at the expense of higher TNPC for all network topologies considered as observed when CPU intensive workloads were considered.

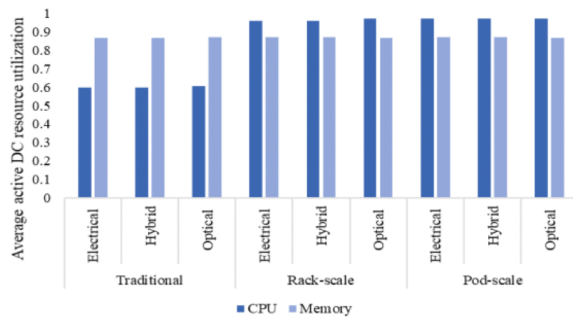

(a) Average active DC resource utilization

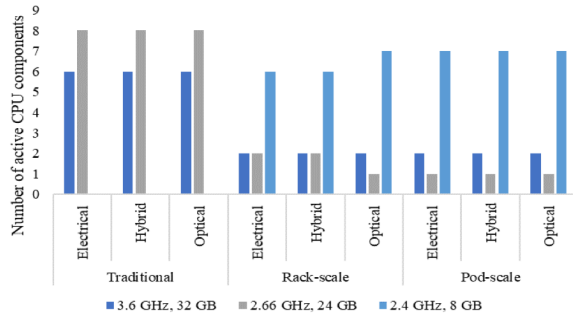

(b) Active CPU under 20 memory intensive workloads

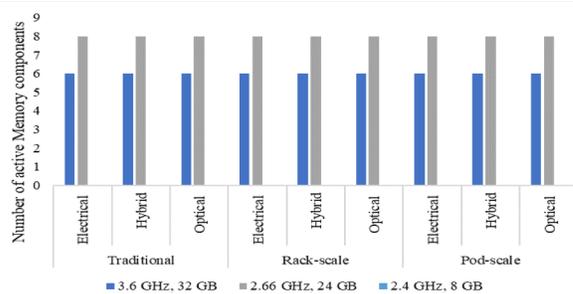

(c) Active memory under 20 memory intensive workloads

Fig. 9. Average utilization and number of active DC resource components under 20 memory intensive workloads.

Relative to similar results obtained when CPU intensive workloads were considered, Fig. 9a shows that memory resource components are highly utilized under all DCs. This is because of the memory intensive nature of the input workloads.

The average utilization of CPU resource components is lower (about 60%) in traditional DCs compared to about 97% average utilization obtained when physical disaggregation of resources at rack-scale and pod-scale are employed. Compared to 84% and 48% average utilization of CPU and memory resource components respectively when CPU intensive workloads are deployed in traditional DCs as shown in Fig. 7a, the average active utilization of CPU and memory components are 60% and 87% respectively when memory intensive workloads are deployed in the same DC as shown in Fig. 9a. Similarly, the average active utilization of CPU and memory resource components are 97% and 87% respectively when memory intensive workloads are deployed in physically disaggregated DCs as seen in Fig. 9a. Relatively, as shown in Fig. 7a, 86% and 93% are the average active resource utilization of CPU and memory resource components respectively when CPU intensive workloads are deployed in physically disaggregated DCs. Hence, greater proportional usage of DC resource components is achieved when memory intensive workloads are provisioned in both traditional and physically disaggregated DCs compared to when CPU intensive workloads are provisioned.

The power consumption profiles of switching components also affects the placement of workload's memory resource demands in rack-scale DC as observed when CPU intensive workloads were provisioned. This is because the trade-off between TMPC and the volume of inter-memory data shuffle traffic remains an important criterion for optimal placement of memory resource demands in rack-scale DCs. Hence, it can be observed that lower TMPC is achieved at the expense of higher network (inter-memory data shuffle) traffic and marginal increase in TNPC when memory intensive workloads are provisioned in a rack-scale DC with optical network topology. In contrast, a rack-scale DC with hybrid or electrical network topology reduces network traffic to achieve lower TNPC by reducing inter-memory traffic in the network as the expense of higher TMPC.

Comparison of TCPC and TMPC obtained in rack-scale and pod-scale DCs under memory intensive workloads also shows that physical disaggregation at rack-scale can achieve equal performance in terms of CPU and memory power efficiencies as physical disaggregation at pod-scale. This is possible (as explained previously) when resource allocation ensures that both CPU and memory resources are available in the appropriate diversity and number in each rack in the rack-scale DC. This requirement is satisfied when 5 memory intensive workloads are provisioned in rack-scale and pod-scale DCs as shown in Fig. 10. These input workloads which require one 3.2 GHz and two 2.4 GHz CPU components and one 32 GB and two 24GB memory components for optimal placement can be adequately provisioned within a single rack of the rack-scale DC to achieve the same optimal TCPC and TMPC observed in the pod-scale DC. Results obtained under 20 memory intensive workloads also reveal the strong impact of capacity constraint during workload placement as shown in Fig. 9a and 9b. For instance, because the memory resource demand of all 20 memory intensive workloads exceed 8 GB, the 8 GB memory components do not have sufficient capacity. Therefore, there is no active 8 GB memory component in the DC as seen in Fig. 9c.

As expected, TNPC of pod-scale DC with electrical/hybrid network topology increases significantly relative to the TNPC of rack-scale DC with electrical/hybrid network topology and the TNPC of the electrical network topology is always higher than that of the hybrid network topology. On the other hand, if optical network topology is adopted in both rack-scale and pod-scale DCs and the number of active racks in both DCs are equal, the TNPC of the rack-scale DC is approximately equal to the TNPC of the pod-scale DC (e.g. under 10, 15 and 20 memory intensive workloads). If the number of active racks in rack-scale DC is less than the same number in pod-scale DC (as is the case when 5 memory intensive workloads are provisioned), then the additional optical switch required per additional active rack is responsible for most of the difference in TNPC as shown in Fig. 10c.

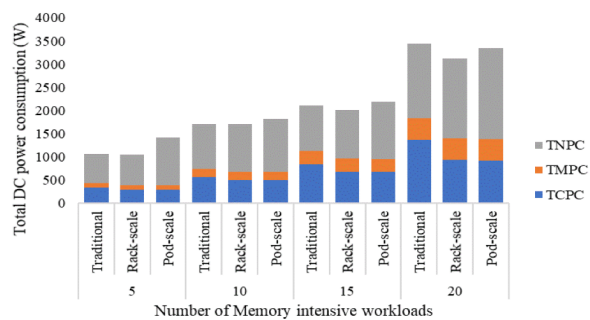

(a) Electrical network topology

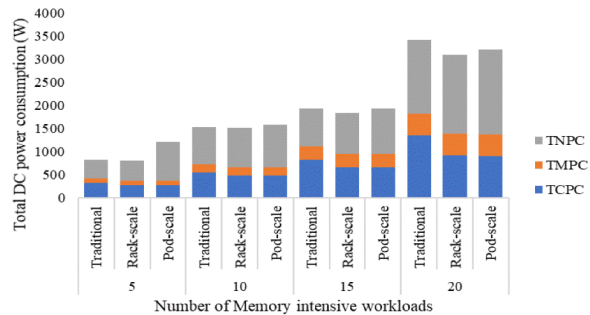

(b) Hybrid network topology

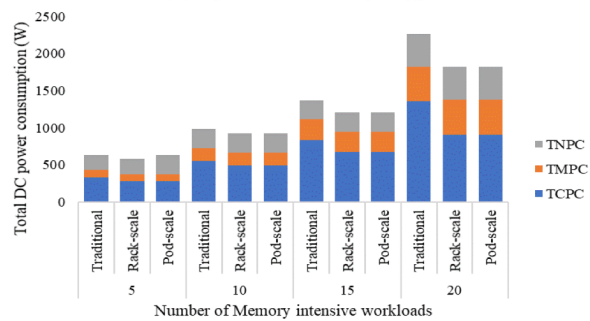

(c) Optical network topology

Fig. 10. Total DC power consumption of composable DCs under memory intensive workloads and different network topologies.

Compared to the traditional DC with optical network topology, rack-scale DC with optical network topology reduces the TDPC by up to 6-20% under varying number of memory intensive workloads as shown in Fig. 10. This is in contrast with 5-8% reduction in TDPC observed when CPU intensive workloads were provisioned in a similar setup. Lower CPU resource demand of memory intensive workloads is responsible for the savings TDPC observed. Relative to the traditional DC with optical network topology, the TDPC of pod-scale DC with optical network topology is also often lower under varying number of memory intensive workloads as shown in Fig. 10c. For example, when 5 memory intensive workloads are optimally provisioned, even though only one rack is active in traditional DC while 2 active racks are required in pod-scale DC, the TDPC in the pod-scale is less than that of traditional DC.

Relative to the TDPC of traditional DC that deploy electrical or hybrid network topology, the TDPC of rack-scale DC with similar network topology is also lower as shown in Fig. 10a and 8b. Likewise, relative to the TDPC of traditional DC that employ hybrid network topology, the TDPC of pod-scale with similar network topology is occasionally (e.g. when 15 and 20 memory intensive workloads are provisioned) lower as shown in Fig. 10b (except when 5 memory intensive workloads are optimally provisioned). Fig. 10a also shows that the TDPC of pod-scale DC with electrical network topology is often (e.g. under 5, 10 and 15 memory intensive workloads) higher than that of traditional DC with a similar network topology because of significant increase in TNPC which exceeds savings in TCPC and TMPC. However, as the number of workloads increases (i.e. 20 memory intensive workloads), the idle power of electrical switches in network topology is increasingly shared and hence their corresponding impact is reduced.

Comparison of TDPC in Fig. 8 to the TDPC in Fig. 10 shows that the TDPC is lower in most scenarios when memory intensive resources are provisioned in corresponding DC type and network. As expected, the TCPC decreases when memory intensive workloads are deployed instead of CPU intensive workloads in all DC and network types while TMPC increases accordingly in all DC and network types. The TNPC is often higher when memory intensive workloads are deployed instead of CPU intensive workloads in corresponding DC and network types. Since the same inter-memory shuffle traffic and inter-resource traffic are adopted when CPU or memory intensive workloads are deployed, high intensity of memory resource demand of input workloads is responsible for this trend. This is because opportunities to consolidate the memory resource demands of workloads that are in the same group into the same memory component or node are reduced when memory intensive workloads are deployed. Hence, more traffic traverses higher network tiers compared to the situation when CPU intensive workloads with finer memory resource demands are deployed in traditional and rack-scale DCs. Relative to traditional and rack-scale DCs, increase in network traffic is minimized in pod-scale DCs because the use of homogenous resourced racks reduces the increase of network traffic as a result of memory resource intensity. Furthermore, the impact of increased network traffic on the TNPC is low when the optical network topology is deployed in DCs compared to when electrical or hybrid topologies are deployed. This is because of electrical switches in both electrical and hybrid network topologies that have a power consumption profile that is proportional to the volume of traffic in the network.

In spite of the marginal benefits enabled by pod-scale physical resource disaggregation over rack-scale resource disaggregation, empirical work by authors in [11] which adopts a disaggregated DC emulator shows that average performance degradation increases when CPU and memory resources are physically disaggregated beyond rack-scale. Moreover, an optical OCS-based network topology whose power consumption profile is agnostic to increase in network traffic as shown in the results above can also guarantee minimal access latency between disaggregated CPU and memory resource via temporal path reservation. Hence, the rack-scale physical resource disaggregation is adopted as the maximum scale of resource disaggregation while optical network topology is also adopted as the default network topology to ensure minimal applications' performance degradation in DCs as the concept of logical resource disaggregation is explored in traditional DCs.

### C. Logical Resource Disaggregation at Rack-Scale

As shown earlier, physical disaggregation of compute resources at rack-scale is sufficient to enable the required flexibility which brings about improvements in resource utilization and overall DC energy efficiency if DC resource allocation ensures that resources are available in appropriate number and/or diversity. It was also observed that the use of an optical network topology within a rack-scale DC ensures maximal network energy efficiency. Furthermore, optical networks are expected enable minimal increase in resource access latency between separated compute resources. On the other hand, optical network topologies are underutilized in traditional DCs where high-bandwidth traffic is node-limited. This negatively impacts the energy efficiency of the optical network topology in a traditional DC. However, features such as high bandwidth and ultra-low latency communication supported by the optical network topology can be effectively used to address underutilization of CPU and memory resources in traditional DCs if the locality constraint present in them is relaxed to permit resource sharing within the rack. This enables a second approach to achieve rack-scale DC via logical resource disaggregation.

In a traditional DC, logical disaggregation implies that CPU and memory components remain in heterogeneous resourced nodes, individual access of each component over a suitable network is allowed and resource locality is confined to each rack in the DC to ensure latency related SLAs are enforced. This is in contrast with a traditional DC where resource locality is confined to each node in the DC. Application of logical disaggregation to the traditional DC can enable performance that approaches that reported under the physically disaggregated rack-scale DC. The performance of logical disaggregation in traditional DC is compared to that of a physically disaggregated rack-scale DC using power consumption as the reference metric. The optical network topology is adopted under both scenarios.

As expected, the performance of the logically disaggregated traditional DC with heterogeneous nodes approaches that of a physically disaggregated rack-scale DC where homogeneous nodes are placed in heterogeneous racks. The results in Fig. 11 show that the TDPC of a logically disaggregated traditional DC is marginally lower than that of a physically disaggregated rack-scale DC for varying number of CPU and memory intensive workloads. The marginal fall in TNPC due to better network utilization efficiency in logically disaggregated traditional DC is responsible for the marginal decrease in TDPC. Co-location of CPU-memory components within the same node in the logically disaggregated traditional DC which occasionally prevents remote memory access over intra-rack network fabric to enables better network energy efficiency is responsible for this. This is not possible in physically disaggregated rack-scale DC. Relative to the physically disaggregated rack-scale DC, the logically disaggregated traditional DC can also guarantee SLAs for ultra-latency sensitive workloads. This is because logically disaggregated traditional DC can default to a traditional DC infrastructure in extreme scenarios (when workloads with very high-sensitivity to increase in memory access latency are being provisioned) to satisfy SLAs while a physically disaggregated rack-scale DC cannot.

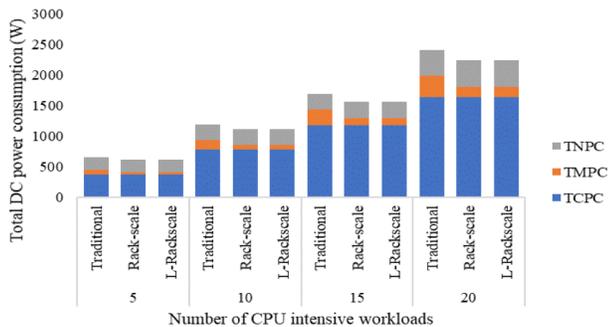

(a) CPU intensive workloads

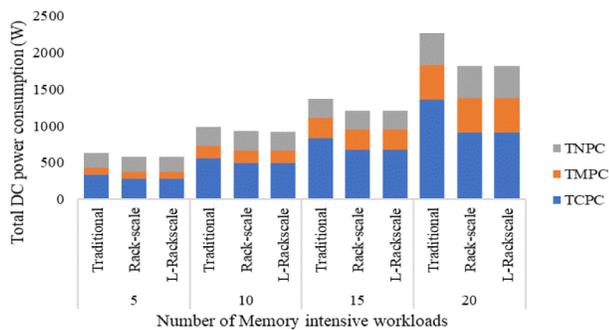

(b) Memory intensive workloads

Fig. 11. Total power consumption of traditional DC and physically and logically disaggregated rack-scale DCs.

Note that to ensure that CPU-memory pairs used to provision some workloads are node-local, placement of CPU and/or memory resource demands in logically disaggregated traditional DC may differ from the placement in physical disaggregated rack-scale DC. Hence, the TCPC and/or TMPC under logically disaggregated traditional DC may marginally increase or decrease relative to the TCPC and/or TMPC under physically disaggregated rack-scale DC. Additionally, if logical disaggregation of traditional DC server is to be adopted, modular resource component design must be adopted in such servers. This will ensure better resource upgrade lifecycle relative to conventional traditional server architecture.

## VI. Micro-Service Architecture in Composable DCs

In the previous sections, monolithic workloads were considered as input workloads for the variety of composable DCs studied. A monolithic workload has fixed resource demands and is designed to run on bespoke physical or virtual hardware for peak performance as seen in previous sections. As an alternative workload architecture, the micro-services workload architecture, proposes the decomposition of monolithic workloads into independent components called micro-services; which perform a specific business function that can be developed, tested, deployed, managed and scaled individually [45]. These related micro-services performing different business functions are thereafter combined together to form an integrated workload which is comparable to the decomposable monolithic workload. For example, a monolithic e-commerce workload which comprise of different intrinsic units performing accounting, inventory and ordering functions can be decoupled into 3 independent micro-services which form a single integrated workload.

Communication between micro-services associated with the same integrated workload is facilitated using well-defined standards or application programming interfaces (APIs) which isolates inner workings of each micro-service [46]–[48]. This novel approach can further enhance scalability, agility, and resource utilization in DCs. In this section, the use of micro-services to create integrated workloads in composable DCs with pre-allocated heterogeneous CPU and memory resources is compared to the use of monolithic workloads.

### A. MILP Model Extension

The model in Section IV.B is extended by introducing sets, parameters, variables, and constraints to establish the relationship between an integrated workload and its micro-services. The additional model set, parameters and variable are given as follows.

**Sets:**

$I$      Set of integrated workloads

**Parameters:**

$CI_i$      CPU capacity of integrated workload $i$

$MI_i$      Memory capacity of integrated workload $i$

$WinI_{wi}$      Indicates the relationship between a micro-service workload $w$ and integrated workload $i$. $WinI_{wi} = 1$ if micro-service workload $w$ is associated with integrated workload $i$. Otherwise, $WinI_{wi} = 0$

**Variable:**

$IS_i$      Indicates the state of integrated workload $m$ i.e. served or unserved. $IS_i = 1$ indicates the integrated workload $i$ is served. Otherwise, $IS_i = 0$

The model in Section IV.B (where set $W$ represents a set of monolithic workloads) continues to represent the DCs supporting monolithic workloads. On the other hand, in DCs supporting micro-services, integrated workloads are represented by set $I$ while set $W$ is the set of micro-services. A group of micro-services make an integrated workload. Based on the results from previous sections of this paper, impact of micro-services is evaluated in a logically disaggregated traditional DC which represents composable DC infrastructure that can support the SLA requirements for both latency sensitive and non-sensitive workloads. The OCS-based optical network topology is adopted; hence, inter-memory and inter-resource communication has little impact on workload resource placement as reported in the previous sections. Since results from the earlier sections also show that inter-memory traffic has limited impact on workload placement over OCS-based optical network topology, the MILP model can be further simplified by excluding variables and parameters required to estimate inter-memory traffic flow.

In addition to constraints $(14) - (31)$, the following constraints establishes the relationship between integrated workloads and micro-services in a composable DC.

$$\sum_{w \in W} \sum_{j \in CR} WC_w WCL_{wj} WinI_{wi} = CI_i IS_i \qquad (36)$$
$$\forall i \in I$$
$$\sum_{w \in W} \sum_{j \in MR} WM_w WML_{wj} WinI_{wi} = MI_i IS_i \qquad (37)$$
$$\forall i \in I$$

Constraints (36) and (37) ensure that an integrated workload is served only if all micro-services associated with the integrated workloads are served. Otherwise, the integrated workload resource demand is rejected. These constraints apply only when micro-services are being provisioned.

Evaluation Scenarios and Results

### B. Evaluation Scenarios and Results

The resource allocation and resource classes described for the traditional DC from earlier sections is maintained. Similarly, the 20 CPU and memory intensive monolithic workloads used in earlier sections are adopted to represent integrated workloads. It is assumed that each integrated workload (CPU or memory intensive) is a unit comprising of two related micro-services that can function independently. Hence, to study the impact of increased workload modularity in composable DCs, each integrated workload is decoupled into two independent micro-service workloads. It is conservatively assumed that the in/out-bound inter-resource communication traffic for each integrated workload is equally shared between its intrinsic micro-services. Table V gives an illustration of all possible evaluation setups for both CPU and memory intensive input workload classes, the acronyms outlined in the Table V are used to represent each setup hereafter.

TABLE V
LIST OF EVALUATION SETUPS

| Setup | Resource allocation | Resource Disaggregation | Workload architecture |
|---|---|---|---|
| TS-Mono | Traditional server | - | Monolithic |
| RS-Mono | Traditional server | Logical | Monolithic |
| TS-Micro | Traditional server | - | Micro-services |
| RS-Micro | Traditional server | Logical | Micro-services |

The impact of different workload architectures and resource disaggregation is studied in DCs using metrics such as DC resource power consumption, number of active DC resources and average active resource utilization as illustrated in Fig. 12 and Fig. 13. The model achieves optimal results via bin-packing of workloads resource demands onto DC resources to achieve optimal resource power efficiency and utilization efficiency within capacity and inter-resource locality constraints.

Note that the placement of workload resource demands, TCPC and TMPC obtained under the both TS-Mono and RS-Mono setups when CPU or memory intensive input workloads are provisioned is like results obtained in earlier sections where similar scenarios have been considered. There are marginal drops in TMPC and TNPC obtained under both TS-Mono and RS-Mono setups relative to values obtained previously. The absence of inter-workload traffic in the revised model is responsible for this observation. Thus, the discussion of results in this Section is focused on the performance of TS-Micro and RS-Micro setups and the comparison of such performance to the results under TS-Mono and RS-Mono setups. To avoid repetition, the results of the TS-Mono and RS-Mono setups are only illustrated in Figures.

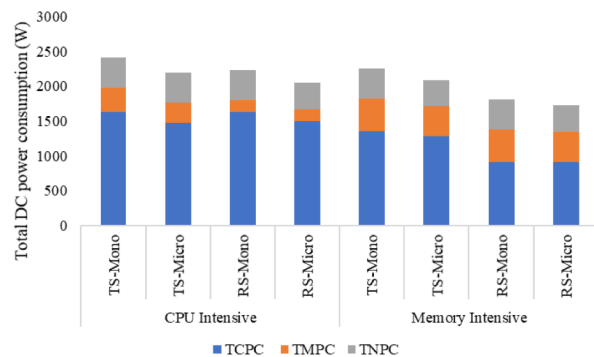

Fig. 12. Total DC power consumption under different setups.

An inspection of the total DC power consumption under the CPU intensive workload in Fig. 12**Error! Reference source not found.** shows that TS-Micro setup reduces the total resource power consumption by 9% compared to the TS-Mono setup. This is because of CPU intensive nature of input workloads and the dominance of CPU resource power consumption over other DC resource types. Reductions in CPU resource power consumption is responsible for more than half (74%) of savings made in total DC power consumption while savings in total memory resource power consumption is responsible for further savings made. On the other hand, total network power consumption remains constant under both setups. A transit from TS-Mono setup to TS-Micro setup under CPU intensive workload leads to 10% and 16% reductions in power consumption of CPUs and memory components respectively. These reductions are enabled via increased workload modularity which leads to improved bin-packing of finer workload resource demands into a different configuration of active servers to achieve greater power efficiency as seen in Fig. 13.

Under the TS-Mono setup, capacity constraint enforced the use of high performance server to provision workloads with CPU resource demand that is above 2.66 GHz. However, the finer resource demands under the TS-Micro setup enable improved bin-packing of demands into servers and consequently ensures improved active resource utilization relative to the TS-Mono setup as shown in Fig. 13a. Nevertheless, it is important to note that CPU intensive nature of input workloads and the strict resource locality constraint of traditional servers led to the high number of active servers with high average CPU utilization and underutilized memory resources. Therefore, disproportionate utilization of DC resources under traditional DC architecture may persist even with increased workload modularity.

A transition to the RS-Mono setup for CPU intensive workloads results in 7% decrease in total DC power consumption compared to the TS-Mono setup. The relaxation of inter-resource locality constraint in logically disaggregated servers is responsible for this observation. Reductions in the power consumption of memory components is solely responsible for the drop observed. The impact of this drop on the total DC power consumption is slightly restricted by the marginal rise in the total network power consumption (as shown in Fig. 14a) resulting from the increase in traffic traversing on-board and intra-rack networks.

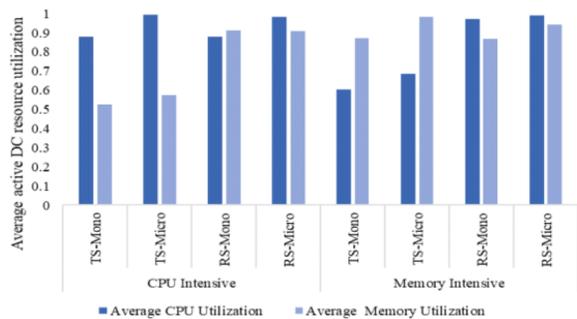

(a) Average active DC resource utilization

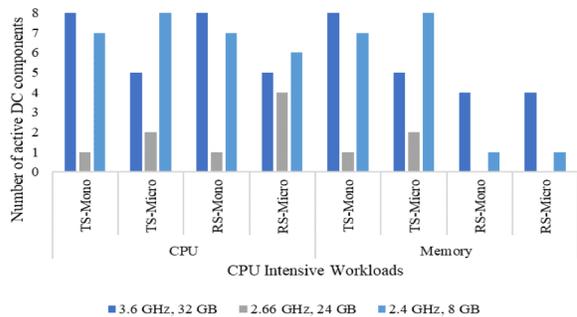

(b) Active DC resources under CPU intensive workloads

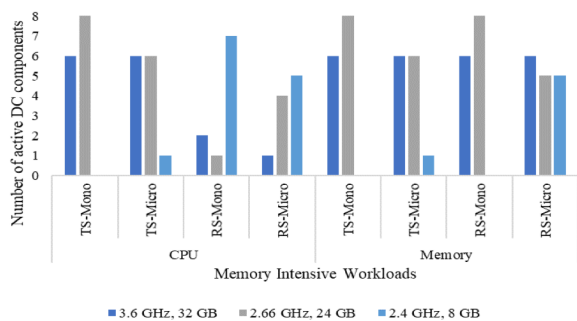

(c) Active DC resources under memory intensive workloads

Fig. 13. Average utilization and number of active DC resource components.

Relative to the TS-Mono setup, logical disaggregation in the RS-Mono setup enabled 51% reduction in TMPC at the expense of 1% rise in TNPC as shown in Fig. 14a. When monolithic workloads are considered under the RS-Mono setup, the same TCPC reported under the TS-Mono setup is obtained despite logical server-disaggregation. This is because improved packing of workloads' intensive CPU resource demands was not feasible. This is sub-optimal, relative to 10% reductions in TCPC achieved under TS-Micro setup as shown in Fig. 14a. Significant reductions in the TMPC under the RS-Mono setup relative to the TS-Mono setup is achieved via consolidation of granular memory resource demands onto reduced number of highly utilized memory components as shown in Fig. 13. CPU components are also highly utilized as shown in Fig. 13a. However, this is because of the CPU intensive nature of workload demands and not because of improved consolidation. Notwithstanding, disaggregation addresses disproportionate usage of DC resources seen in traditional DC infrastructure. This is because greater resource modularity enabled by disaggregation increases flexibility when CPU and memory components used to provision workload resource demands are selected. Hence, only the minimum number of resources needed to effectively satisfy each type of workload resource demand are activated.

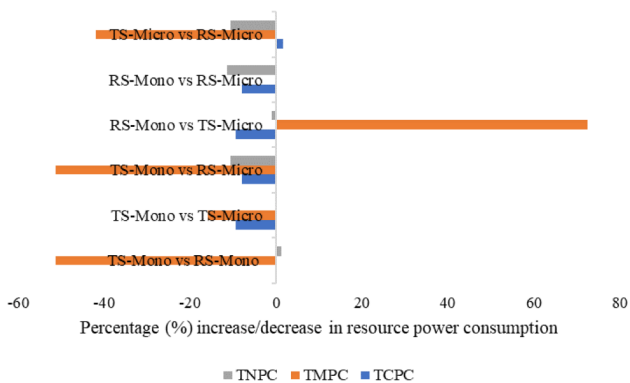

(a) CPU intensive

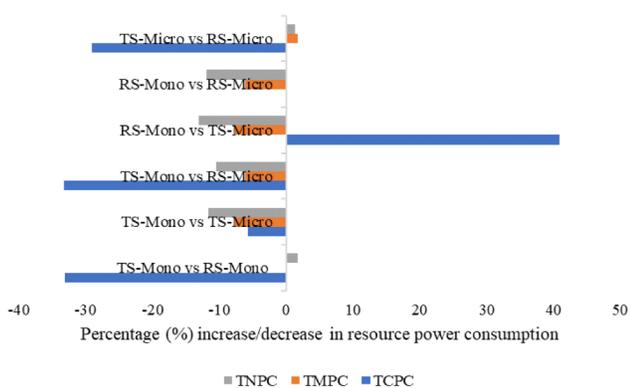

(b) Memory intensive

Fig. 14. Percentage decrease in DC resource power consumption.

Adoption of both logical disaggregation and micro-service architecture under the RS-Micro setup addresses the limitation of RS-Mono setup to achieve additional savings in total DC power consumption. Relative to TS-Mono, TS-Micro and RS-Mono setups, RS-Micro leads to 15%, 7% and 8% reduction in total DC power consumption respectively. Thus, the RS-Micro setup activates unattained potentials of TS-Mono, TS-Micro and RS-Mono setups for optimal DC efficiency via improved bin-packing of fine resource demands onto active resource components. For example, relative to the RS-Mono setup, a move to the RS-micro leads to 8% reduction in TCPC. This power savings is achieved via the use of a different configuration of active CPU components as shown in Fig. 13a. At the same time, only marginal reduction in total memory power consumption achieved when RS-Micro setup is adopted over the RS-Mono setup. This is because the granularity of non-intensive memory resource demands is sufficient to achieve optimal memory power consumption in a disaggregated DC without adoption of the micro-service architecture.

Similar observations to those reported above for CPU intensive workload class are repeated under memory intensive workload classes. Hence, 7%, 20% and 23% reduction in the total DC power consumption of the TS-Mono setup was achieved via the deployment of TS-Micro, RS-Mono and RS-Micro setups respectively for the memory intensive workload class.

The TS-Micro setup enables reductions in CPU, memory and network power consumption compared to the TS-Mono setup for memory intensive workloads as shown in Fig. 14b. When workloads are memory intensive, the TS-Micro setup is more power efficient for provisioning memory resource demands than the RS-Mono setup. On the other hand, the RS-Mono setup is more power efficient for provisioning CPU resource demand of memory intensive workloads compared to the TS-Micro setup as shown in Fig. 14b. This is because the memory intensive nature of monolithic workloads limits optimal consolidation of such workload memory resource demand in the RS-Mono setup. Thus, resulting into higher memory power consumption under RS-Mono setup relative to the TS-Micro setup as shown in Fig. 12 and Fig. 14b. However, reductions in the TCPC which significantly dominates memory resource utilization inefficiencies leads to lower TDCPC in RS-Mono setup relative to the TS-Micro setup when memory intensive monolithic workloads are provisioned. This is because of the dominance of CPU resource power consumption over memory resources. The RS-Micro setup addresses the limitation of the RS-Mono setup. Relative to the TS-Micro, RS-Micro can deliver equal or better TCPC and TMPC. Hence, surpassing the limitations of the RS-Mono setup as shown in Fig. 12.

When the class of input workloads is neither CPU intensive nor memory intensive, we expected limited reduction in the total DC power consumption of the TS-Mono setup relative to TS-Micro, RS-Mono or RS-Micro setup. This is because the impact of capacity constraint is relaxed when workloads are not CPU intensive nor memory intensive. Likewise, lower resource intensity of such workloads implies that they are likely to approach optimal performance (i.e. energy efficiency) under the TS-Mono setup. Hence, only marginal performance gains will be achieved if such workloads are provisioned under TS-Micro, RS-Mono or RS-Micro setup.

## VII. HEURISTIC FOR ENERGY EFFICIENT PLACEMENT OF WORKLOADS IN COMPOSABLE DCs

Results from the MILP model showed that efficient placement of monolithic and micro-service workloads in composable DCs is required to reach optimal energy efficiency in DCs with heterogeneous resource types. This task which was formulated as MILP model is a multi-dimensional bin packing problem with multi-sized bins of different resource types which may or may not be co-located. The classical one-dimensional bin-packing problem is NP-complete, hence multi-dimensional bin packing problem with bins which have varying capacity and power consumption is also NP-complete. Therefore, only approximation algorithms that mimic the results of the MILP model for different workload classes can provide best-fit solutions without the need for exhaustive search and permutation of workloads and DC resources. To this end, a heuristic for energy efficient placement (HEEP) of workloads in a rack-scale composable DCs is proposed.

### A. HEEP Algorithm Description

HEEP is a greedy algorithm designed to be deployed with a centralized or distributed DC orchestration and control platform. It is required that the orchestration platform maintain global knowledge of resource state, utilization and power consumption across the DC as in software defined DCs. The orchestrator should also be aware of input workload resource demands over a specific time frame. The orchestrator is responsible for energy efficient placement of different workload classes in different DC architectures. The flow chart of the HEEP algorithm is given in Fig. 15.

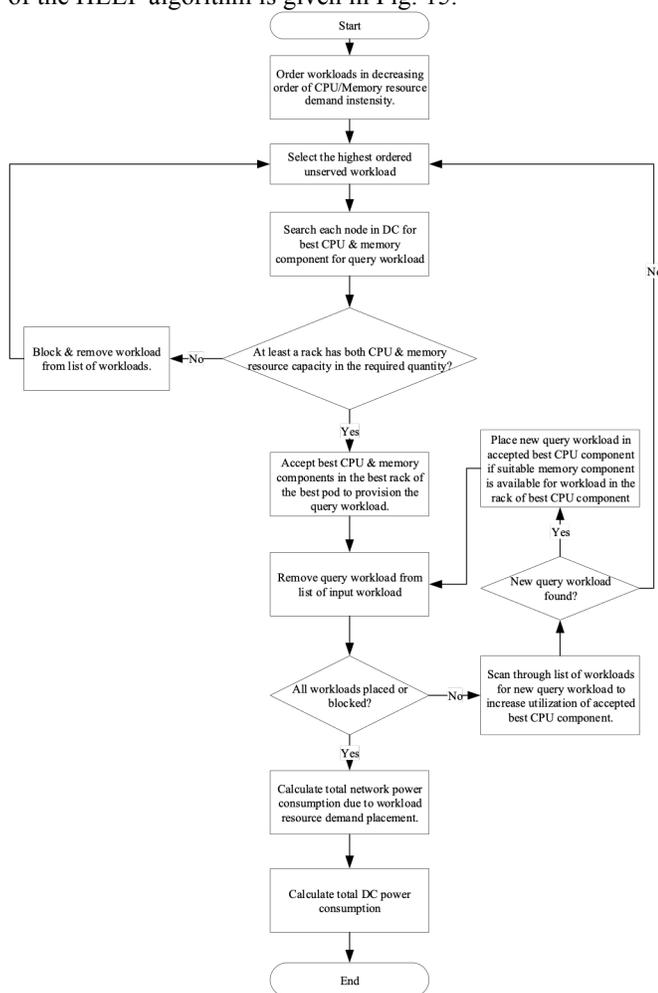

Fig. 15. HEEP algorithm flow chart.

Given a set of input workloads with CPU or memory resources demands with inter-resource communication traffic, the algorithm achieves minimal total DC resource power consumption through the following steps. Input workloads are arranged in descending order of CPU/memory resource demands with respect to the workload class under consideration to minimize blocking of workloads with ultra-high resource demands since the heuristic adopts a greedy approach. The sorted list of workloads is the input to the algorithm and is called the job list hereafter. The highest ordered workload in job list is set as the default query workload. The central orchestrator adopts a divide and conquer approach by searching all nodes across the DC for the best CPU and memory resource component within each node to provision the query workload. If feasible, each node returns a candidate CPU and/or memory component to provision the query workload alongside the corresponding power consumption and utilization of such component in the candidate node.

Note that if a rack with sufficient CPU and memory resource capacities to host the query workload does not exist (i.e. the workload blocking criterion) the query workload is blocked and removed from the job list. Otherwise, the central orchestrator selects the best CPU and memory components to provision the query workload in each rack using pre-defined utilization thresholds for CPU and memory components as illustrated in Fig. 16. Afterwards, the central orchestrator also selects the best rack to provision the query workload in each pod based on the utilization threshold of the best candidate CPU component of each rack. Finally, the central orchestrator selects the best pod to provision the query workload also based on the utilization threshold of the best candidate CPU component in the best candidate rack of each pod. Afterwards, the query workload is removed from the job list. These steps imply that the algorithm searches all nodes, racks and pods to obtain the best candidate resource components to host each workload resource demand.

If the job list is not empty, the heuristic attempts to select the next query workload by scanning through the job list for a workload that will fit into unused capacities of active resources in the present best rack to increase utilization. For resource intensive workloads, the scan attempts to fill the idle CPU capacity of the present best CPU component. This is because CPU components have higher peak power consumption compared to memory components. If the input workload class is CPU or memory intensive, the scan orders workloads in descending order of CPU resource demand intensity, thus giving higher preference to CPU resource demand. Higher preference is also given to CPU resource demands when memory intensive workloads are being considered because it ensures CPU biased resource allocation which leads to optimal total DC resource power consumption.

A successful scan returns a workload which becomes the new query workload that is placed in the present best rack. Otherwise, the highest ordered unserved workload in the job list is selected as the new query workload. When all workloads have been successfully placed or blocked, the algorithm estimates the total network power consumption resulting from workloads' CPU and memory resource demand placement across the DC, and subsequently reports the total DC resource power consumption before stopping. Note that in the algorithm, ties are always broken by selecting the first item (i.e. workload, resource component or rack) that appears.

Given a list of candidate components for a specific resource demand type, their component class and the resulting utilization of placing the resource demand in the candidate components, Fig. 16 gives an illustration of the steps taken to select the best components based on relative utilization thresholds of each component class. A component in the most energy efficient component class is in the list of candidate components is given priority. However, such component is only adopted as the best component under the following conditions:

- If the corresponding utilization ($U$) after the placement of the query resource demand is less than or equal to the lower utilization threshold ($\lambda_k$) defined for that component's class $k$.
- If the corresponding utilization ($U$) after the placement of the query resource demand is greater than the upper

utilization threshold ($\Lambda_k$) defined for that component's class $k$.

- If the component is the last (and/or only) component in the list of candidate component being evaluated to support the query resource demand.

Otherwise, the most energy efficient component is removed from the list of candidate components for the query resource demand and a new component is selected from the class of the most energy efficient component in list of candidate components. The lower and upper utilization threshold must be defined based on the relative energy efficiency of classes of resource components deployed in the composable DC.

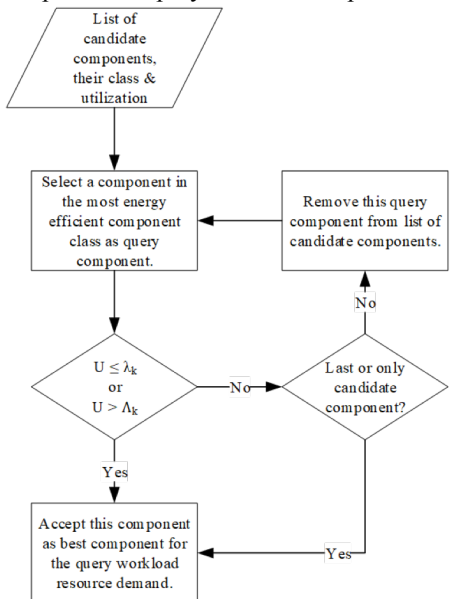

Fig. 16. Best component selection based on component's class utilization.

Using the CPU or memory resource component classes listed in Table I, the lower and upper utilization thresholds for resource component classes are defined as follows.

- Given a set of candidate component classes $K$ for a given resource type which is sorted in descending order energy efficiency.

- The lower utilization threshold ($\lambda_k$) of all component classes (except the last component class) in the set is always 0.5. This is because utilizing 50% (or less) of a component class's capacity to provision the most intensive resource demand leaves half (or more) of the class's capacity in an idle state. This idle capacity can be utilized to provision another intensive resource demand.

- The upper utilization threshold ($\Lambda_k$) of all component classes (except last component) in the set is defined by (38). This is because to energy efficiently utilize a resource component of class $k$ which has higher capacity and power consumption than another resource component of class $k + 1$, the resource demanded by all workloads placed in the component with greater power efficiency must be greater than the capacity of the component with lower power efficiency.

Consequently, if it is assumed that all CPU or memory resource component classes listed in Table I are present in a list of candidate components for a given query workload, the relative utilization thresholds of all candidate component classes (except last component) are as given in Table VI.

$$\Lambda_k = \frac{Capacity_{k+1}}{Capacity_k} \tag{38}$$
$$\forall k \in K : k \neq |K|$$

TABLE VI
RELATIVE UTILIZATION THRESHOLD FOR COMPONENT CLASSES

| CPU components | | | |
|---|---|---|---|
| $k \in K$ | Capacity | $\lambda_k$ | $\Lambda_k$ |
| 1 | 3.6 GHz | 0.5 | 0.73889 |
| 2 | 2.66 GHz | 0.5 | 0.90226 |
| 3 | 2.4 GHz | N/A | N/A |

| Memory components | | | |
|---|---|---|---|
| $k \in K$ | Capacity | $\lambda_k$ | $\Lambda_k$ |
| 1 | 32 GB | 0.5 | 0.75 |
| 2 | 24 GB | 0.5 | 0.333 |
| 3 | 8 GB | N/A | N/A |

It is important to note that the HEEP algorithm was designed for rack-scale composable DCs based on the discussions in earlier sections of this paper. However, the heuristic can be extended to support a pod-scale composable DC should the need arise. This can be achieved by revising the workload blocking criterion in the heuristic to consider availability of all suitable resource component types at the pod-level rather than at the rack-level as described above.

### B. Performance Evaluation

The performance of the HEEP algorithm is evaluated via a comparison with the optimal results obtained by solving the MILP model in preceding sections using similar input parameters. The results show that the total DC resource power consumption achieved by the HEEP algorithm replicates the trend reported from solving the MILP under different scenarios. Fig. 17 shows the results obtained from MILP model and those obtained from the HEEP algorithm when 20 CPU and memory intensive are provisioned under different workload and DC architectures.

Similar to results obtained by solving the MILP model, the HEEP algorithm also shows that the highest total DC resource power consumption is observed when monolithic workloads are deployed in traditional DCs. Results obtained using the HEEP algorithm also shows the effectiveness of logical server disaggregation and adoption of more granular workload architecture towards the reduction of total DC power consumption. Relative to the MILP model results, Fig.17a shows that the highest percentage increase in the total DC resource power consumption observed when HEEP algorithm is deployed for CPU intensive workloads under different DC and workload architectures is 14%. Under 20 memory intensive workloads, percentage increase in total DC resource power consumption of the HEEP algorithm relative to the results obtained from solving the MILP model does not exceed 11% for all scenarios in Fig. 17b.

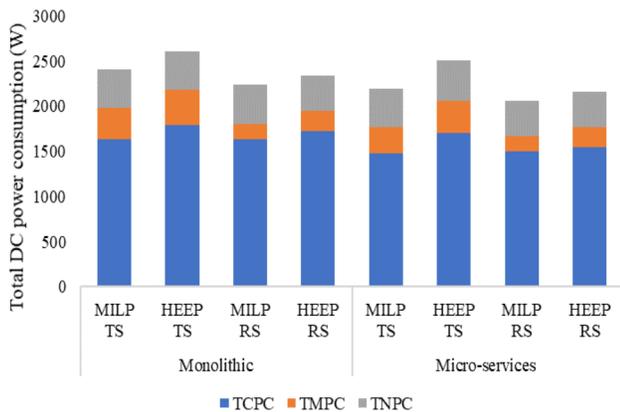

(a) CPU intensive

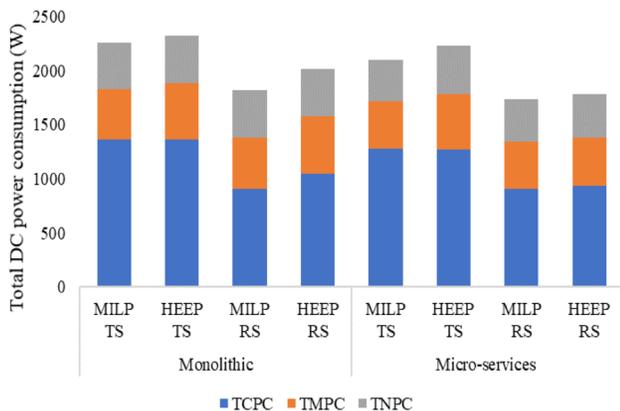

(b) Memory intensive

Fig. 17. HEEP and MILP model power consumption under 20 workloads.

Fig. 18 shows the total DC power consumption when varying number of CPU and memory intensive workloads are used as input to both the MILP model and the HEEP algorithm. For the small number of in workloads (i.e. 5 - 20) considered, the results show that the HEEP algorithm is more effective for provisioning workloads in composable DCs than it is for traditional DCs. Relative to the MILP model, the average percentage increase in the total DC power consumption when the HEEP algorithm is adopted in composable DCs are 6% and 8% for CPU and memory intensive workloads classes respectively. Compared to the MILP model, these values increase to 11% and 19% for CPU and memory intensive workload classes respectively when the HEEP algorithm is adopted in the traditional DC. It is expected that the margins between results obtained using the MILP model and HEEP algorithm will decrease as the number and diversity of input workload demands in each workload class increase. This is because of the expected increase in the probability of workload consolidation onto DC resources. This will in turn lead to total DC resource power consumption that approach those reported from solving the MILP model.

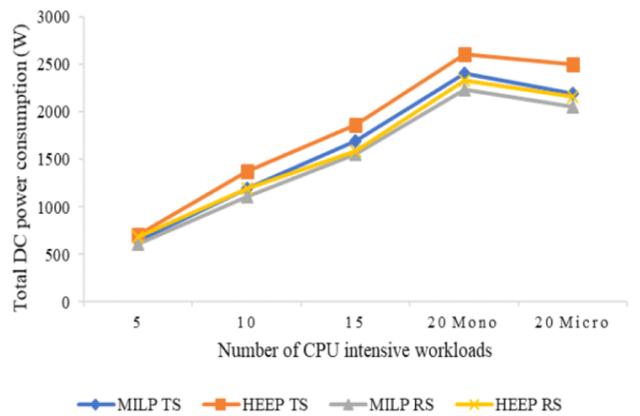

(a) CPU intensive

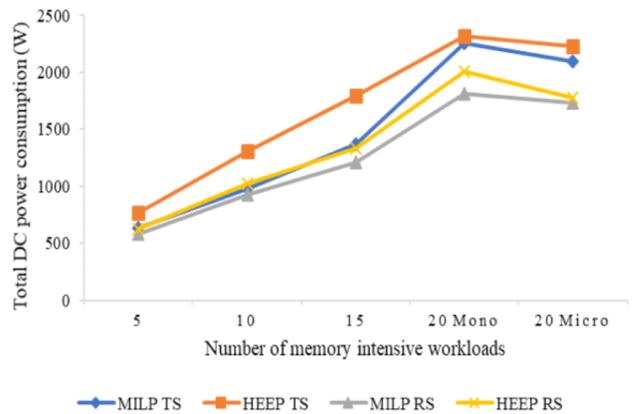

(b) Memory intensive

Fig. 18. Comparison of HEEP and MILP model power consumption.

## VIII. Conclusions

In this paper we formulated a MILP model to evaluate the performance of physical disaggregation of compute resources at rack-scale and pod-scale under selected electrical, optical hybrid network topologies. Relative to resource disaggregation at pod-scale in composable DCs, our results showed that physical disaggregation of compute resources at rack-scale is sufficient to achieve optimal resource utilization efficiency provided that appropriate distribution of resource (both in number and/or diversity) is ensured during resource allocation. Adoption of optical network topology in composable DCs ensures optimal overall DC energy efficiency. Physical resource disaggregation of traditional DC servers at rack-scale leads to better (6-20%) savings in overall power consumption when memory intensive workloads are provisioned compared to 5-8% savings in total DC power consumption obtained when CPU intensive workloads are provisioned. Relative to the physical resource disaggregation at rack-scale, logical disaggregation of traditional servers over an optical network topology within a rack ensures a composable infrastructure suitable for all workload categories (i.e. workloads with low-sensitivity or high-sensitivity to increase in memory access latency) while ensuring improved network power efficiency.

The paper also explored the impact of micro-service architecture on overall DC operational and energy efficiency in both traditional and composable DCs. The results show that in spite of improvements in total resource power consumption and resource utilization in traditional DC enabled by increased workload modularity, disproportionate utilization of DC resources may persist. While disaggregation addresses this limitation of traditional DCs, utilization of disaggregated DCs is limited when resource intensive monolithic workloads are deployed. Ultimately, a combination of composable infrastructure and increased workload modularity enables optimal resource utilization and energy efficiencies in DCs. Hence, these both approaches are complementary and DC operators should leverage on the strengths of both approaches to enable optimal DC resource utilization and energy efficiency.

Finally, we proposed a real-time heuristic for energy efficient placement of workloads in composable DCs. The total DC power consumption obtained using the HEEP algorithm approached the exact results obtained by solving the MILP model under CPU intensive and memory intensive workload classes. Future work will explore the impact of adopting composable DCs for real-time applications in fog computing era which prefer computation at the edge of telecom networks. In a more practical setup, a study of the adoption of capacitated optical networks in composable DCs is also planned.

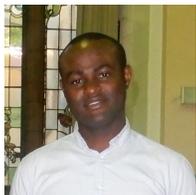

**Opeyemi O. Ajibola** received the B.sc. degree (High Hons.) in Electrical and Electronic Engineering from Eastern Mediterranean University, Famagusta, North – Cyprus, in 2011 and the M.Sc. degree (with distinction) in Digital Communications Networks from University of Leeds, Leeds, UK in 2015. He is currently working towards the PhD degree in the School of Electronic and Electrical Engineering, University of Leeds, Leeds, UK. From 2012 to 2013, he was a Wireless Solution Sales Engineer with Huawei Technologies, Abuja, Nigeria. In 2014, he joined Federal University Oye-Ekiti, Ekiti State, Nigeria as a graduate assistant. His research interests include composable datacenter infrastructures, energy efficient datacenter and communication networks, energy efficient cloud and fog/edge computing and the Internet of Things.

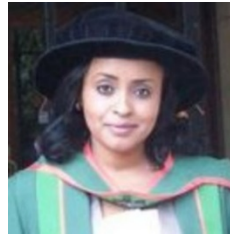

**Taisir E. H. El-Gorashi** received the B.S. degree (first-class Hons.) in Electrical and Electronic Engineering from the University of Khartoum, Khartoum, Sudan, in 2004, the M.Sc. degree (with distinction) in Photonic and Communication Systems from the University of Wales, Swansea, UK, in 2005, and the PhD degree in Optical Networking from the University of Leeds, Leeds, UK, in 2010. She is currently a Lecturer in optical networks in the School of Electronic and Electrical Engineering, University of Leeds. Previously, she held a Postdoctoral Research post at the University of Leeds (2010– 2014), where she focused on the energy efficiency of optical networks investigating the use of renewable energy in core networks, green IP over WDM networks with datacenters, energy efficient physical topology design, energy efficiency of content distribution networks, distributed cloud computing, network virtualization and big data. In 2012, she was a BT Research Fellow, where she developed energy efficient hybrid wireless-optical broadband access networks and explored the dynamics of TV viewing behavior and program popularity. The energy efficiency techniques developed during her postdoctoral research contributed 3 out of the 8 carefully chosen core network energy efficiency improvement measures recommended by the GreenTouch consortium for every operator network worldwide. Her work led to several invited talks at GreenTouch, Bell Labs, Optical Network Design and Modelling conference, Optical Fiber Communications conference, International Conference on Computer Communications, EU Future Internet Assembly, IEEE Sustainable ICT Summit and IEEE 5G World Forum and collaboration with Nokia and Huawei.

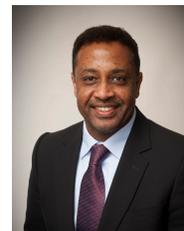

**Professor Jaafar Elmirghani** is Fellow of IEEE, Fellow of the IET, Fellow of the Institute of Physics and is the Director of the Institute of Communication and Power Networks and Professor of Communication Networks and Systems within the School of Electronic and Electrical Engineering, University of Leeds, UK. He joined Leeds in 2007 having been full professor and chair in Optical Communications at the University of Wales Swansea 2000-2007.

He received the BSc in Electrical Engineering, First Class Honours from the University of Khartoum in 1989 and was awarded all 4 prizes in the department for academic distinction. He received the PhD in the synchronization of optical systems and optical receiver design from the University of Huddersfield UK in 1994 and the DSc in Communication Systems and Networks from University of Leeds, UK, in 2012. He co-authored Photonic Switching Technology: Systems and Networks, (Wiley) and has published over 550 papers.

He was Chairman of the IEEE UK and RI Communications Chapter and was Chairman of IEEE Comsoc Transmission Access and Optical Systems Committee and Chairman of IEEE Comsoc Signal Processing and Communication Electronics (SPCE) Committee. He was a member of IEEE ComSoc Technical Activities Council' (TAC), is an editor of IEEE Communications Magazine and has been on the technical program committee of 41 IEEE ICC/GLOBECOM conferences between 1995 and 2020 including 19 times as Symposium Chair. He was founding Chair of the Advanced Signal Processing for Communication Symposium which started at IEEE GLOBECOM'99 and has continued since at every ICC and GLOBECOM. Prof. Elmirghani was also founding Chair of the first IEEE ICC/GLOBECOM optical symposium at GLOBECOM'00, the Future Photonic Network Technologies, Architectures and Protocols Symposium. He chaired this Symposium, which continues to date. He was the founding chair of the first Green Track at ICC/GLOBECOM at GLOBECOM 2011, and is Chair of the IEEE Sustainable ICT Initiative, a pan IEEE Societies Initiative responsible for Green ICT activities across IEEE, 2012-present. He has given over 90 invited and keynote talks over the past 15 years.

He received the IEEE Communications Society 2005 Hal Sobol award for exemplary service to meetings and conferences, the IEEE Communications Society 2005 Chapter Achievement award, the University of Wales Swansea inaugural 'Outstanding Research Achievement Award', 2006, the IEEE Communications Society Signal Processing and Communication Electronics outstanding service award, 2009, a best paper award at IEEE ICC'2013, the IEEE Comsoc Transmission Access and Optical Systems outstanding Service award 2015 in recognition of "Leadership and Contributions to the Area of Green Communications", the GreenTouch 1000x award in 2015 for "pioneering research contributions to the field of energy efficiency in telecommunications", the IET 2016 Premium Award for best paper in IET Optoelectronics, shared the 2016 Edison Award in the collective disruption category with a team of 6 from GreenTouch for their joint work on the GreenMeter, the IEEE Communications Society Transmission, Access and Optical Systems technical committee 2020 Outstanding Technical Achievement Award for outstanding contributions to the "energy efficiency of optical communication systems and networks". He was named among the top 2% of scientists in the world by citations in 2019 in Elsevier Scopus, Stanford University database which includes the top 2% of scientists in 22 scientific disciplines and 176 sub-domains. He was elected Fellow of IEEE for "Contributions to Energy-Efficient Communications," 2021.

He is currently an Area Editor of IEEE Journal on Selected Areas in Communications series on Machine Learning for Communications, an editor of IEEE Journal of Lightwave Technology, IET Optoelectronics and Journal of Optical Communications, and was editor of IEEE Communications Surveys and Tutorials and IEEE Journal on Selected Areas in Communications series on Green Communications and Networking. He was Co-Chair of the GreenTouch Wired, Core and Access Networks Working Group, an adviser to the Commonwealth Scholarship Commission, member of the Royal Society International Joint Projects Panel and member of the Engineering and Physical Sciences Research Council (EPSRC) College.

He has been awarded in excess of £30 million in grants to date from EPSRC, the EU and industry and has held prestigious fellowships funded by the Royal Society and by BT. He was an IEEE Comsoc Distinguished Lecturer 2013-2016. He was PI of the £6m EPSRC Intelligent Energy Aware Networks (INTERNET) Programme Grant, 2010-2016 and is currently PI of the EPSRC £6.6m Terabit Bidirectional Multi-user Optical Wireless System (TOWS) for 6G LiFi, 2019-2024. He leads a number of research projects and has research interests in communication networks, wireless and optical communication systems.